\documentclass{article}
\font\sqi=cmssq8
\def\DR{\rm I\kern-1.45pt\rm R}
\def\DC{\kern2pt {\hbox{\sqi I}}\kern-4.2pt\rm C}
\def\DH{\rm I\kern-1.5pt\rm H\kern-1.5pt\rm I}

\newcommand{\ben}{\begin{enumerate}}
\newcommand{\een}{\end{enumerate}}
\newcommand{\beq}{\begin{equation}}
\newcommand{\eeq}{\end{equation}}
\newcommand{\bse}{\begin{subequation}}
\newcommand{\ese}{\end{subequation}}
\newcommand{\bea}{\begin{eqnarray}}
\newcommand{\eea}{\end{eqnarray}}
\newcommand{\bc}{\begin{center}}
\newcommand{\ec}{\end{center}}
\newcommand{\bQ}{{\overline Q}}

\newcommand{\bs}{\mbox{\boldmath $\sigma$}}
\newcommand{\bg}{\mbox{\boldmath $\gamma$}}
\newcommand{\bp}{\mbox{\boldmath $\pi$}}

\newcommand{\ch}{{\tt h}}

\begin{document}
\begin{center}
{\large\bf Elements of (super-)Hamiltonian formalism}
\\[2mm]

{\large Armen Nersessian}\\[2mm]

{\it Artsakh State University, Stepanakert \&
Yerevan State University, Yerevan,
ARMENIA}\end{center}
\begin{abstract}
In these lectures we discuss some basic aspects of Hamiltonian formalism,
which  usually do not appear in standard texbooks on classical mechanics for physicists.
We pay special attention  to the procedure of Hamiltonian reduction
illustrating it by the
examples related to  Hopf maps.
Then we  briefly discuss  the supergeneralisation(s)
of the Hamiltonian formalism and present
some simple models of supersymmetric mechanics on K\"ahler manifolds.
\end{abstract}
\setcounter{equation}{0}
\section*{Introduction}
The goal of these lectures is to convince the reader to construct the
supersymmetric mechanics within the Hamiltonian framework, or, at least,
 to combine the superfield approach with the existing methods of Hamiltonian mechanics.
The standard approach to construct the supersymmetric mechanics
with more than two supercharges is the Lagrangian superfield approach.
Surely, superfield formalism is a quite powerful method for the construction
of supersymmetric theories.
However, all superfield formalisms, being developed {\sl \'a priori}
for field theory, are convenient
for the construction of the field-theoretical models, which are covariant
with respect to space-time coordinate transformations.
However, the supermultiplets (i.e. the basic ingredients of superfield formalisms)
do not respect the transformations mixing field variables.
On the other hand, in supersymmetric mechanics these variables appear as  spatial
coordinates. In other words, the superfield approach, being applied to
supersymmetric mechanics, provides us with a local construction
of mechanical models.
Moreover, the obtained models need
to be re-formulated in the Hamiltonian framework, for the subsequent quantization.
In addition, many of the numerous methods and statements
in the Hamiltonian formalism could be easily extended to supersymmetric systems and applied there.
Independently from the specific preferences, the ``Hamiltonian view" of the
existing models of supersymmetric mechanics, which were
 built within the superfield approach,
 could establish unexpected links between different supermultiplets and models.
Finally, the superfield methods seem to be too general
in the context of simple mechanical systems.

For this reason, we tried to present some
elements of Hamiltonian formalism,  which do not usually
appear in the standard textbooks on classical mechanics, but appear to be
useful  in the context of supersymmetric mechanics.
We pay much attention  to the procedure of Hamiltonian reduction, having in mind
that it could be used for the construction of the lower-dimensional
supersymmetric models from the existing higher-dimensional ones.
Also, we devote a special   attention to the Hopf maps and  K\"ahler spaces, which are
 typical structures in supersymmetric systems.
Indeed, to
extend the number of supersymmetries
(without extension of the fermionic degrees of freedom)
we usually equip the configuration/phase  space  with complex structures
and restrict them to be K\"ahler, hyper-K\"ahler, quaternionic
and so on, often via a choice of the appropriate
supermultiplets related to the real, complex, quaternionic structures.
We illustrated  these matters by examples of Hamiltonian reductions
related with Hopf maps, having in mind that themy could be straightforwardly applied to
supersymmetric systems.
Also, we included some less known material related with Hopf fibrations.
It concerns the generalization  of the oscillator to spheres, complex projective spaces,
and quaternionic projective spaces, as well as the reduction
of the oscillator systems to Coulomb ones.

Most of the presented constructions are developed only
for the zero and first Hopf maps. We tried to present them in the way,
 which will clearly show, how to extend them to the second Hopf map
 and the quaternionic case.

The last two sections  are
 devoted to the super-Hamiltonian formalism.
We present the superextensions of the Hamiltonian constructions,
 underlying the specific ``super"-properties, and present some examples.
Then we provide the list of supersymmetric mechanics
constructed within the Hamiltonian approach.
Also in this case, we tried to arrange the material in such a way,
as to make clear the relation of these constructions to complex structures
and their possible extension to quaternionic ones.

The main references to the generic facts about Hamiltonian mechanics are the
excellent textbooks \cite{arnold,perelomov}, and  on the supergeometry there exist the
 monographs \cite{berezin,voronov}.
 There are numerous reviews on supersymmetric mechanics. In our opinion
 the best introduction to the subject is given in refs. \cite{krive,sukh}.

\setcounter{equation}{0}
\section{Hamiltonian formalism}
In this Section we present some basic facts about the Hamiltonian
formalism, which could be straightforwardly extended
to the super-Hamiltonian systems.

We restrict ourselves to considering Hamiltonian systems with nondegenerate
Poisson brackets. These brackets are defined, locally, by the expressions
\beq
\{f,g\}=\frac{\partial f}{\partial x^i}
\omega^{ij}(x)\frac{\partial g}{\partial x^j}, \qquad \det \omega^{ij}\neq 0,
\label{1}\eeq
where
\bea
\label{2}&\{f,g\}=-\{g,f\},\quad \Leftrightarrow\quad \omega^{ij}=-\omega^{ji}&\\
\label{3}& \{\{f,g\},h\}+ {\rm cycl.perm} (f,g,h)=0,\quad
\Leftrightarrow\quad \omega^{ij}_{,n}\omega^{nk}+{\rm cycl.perm } (i,j,k)\;=0.&
\eea
The Eq. (\ref{2})  is known as a ``antisymmetricity condition",
and the Eq.(\ref{3}) is called Jacobi identity.
Owing to the nondegeneracy of the matrix $\omega^{ij}$, one can construct the
nondegenerate
two-form, which is closed due to Jacobi identity
\beq
\omega=\frac 12 \omega_{ij}dx^i\wedge dx^j\;\;: \;\;
d\omega=0\;\Leftrightarrow \;\omega_{ij,k} +{\rm cycl.perm } (i,j,k)=0.
\label{4}\eeq
The manifold $M$ equipped with such a form, is called symplectic manifold,
and denoted by $(M,\omega)$.
It is clear that $M$ is an even-dimensional manifold, $dim M=2N$.\\
The Hamiltonian system is defined by the triple $(M,\omega, H)$, where
$H(x)$ is a scalar function called Hamiltonian.

The Hamitonian equations of motion yield the vector field
preserving the symplectic form $\omega$
\beq
\frac{dx^i}{dt}=\{H,x^i\}={ V}^i_{H} \;:\qquad
{\cal L}_{V_{H}}\omega=0.
\label{5}\eeq
Here ${\cal L}_{{\bf V}}$ denotes the Lie derivative along vector field ${\bf V}$.

Vice versa, any vector field, preserving
the symplectic structure, is locally a Hamiltonian one. The easiest
 way to see it
is to  use   homotopy formula
\beq
\imath_{{\bf V}}d\omega+d\imath_{{\bf V}}\omega=
{\cal L}_{{\bf V}}\omega \;\Rightarrow \; d\imath_{{\bf V}}\omega=0.
\label{6}\eeq
Hence, $\imath_{{\bf V}}\omega$ is a closed one-form and could be locally
presented as follows: $\imath_{{\bf V}}\omega=dH(x)$.
The local function $H(x)$ is precisely the Hamiltonian, generating
 the vector field ${\bf V}$.
The transformations preserving the symplectic structure are called
{\sl canonical transformations}.

Any symplectic structure could be locally presented in the form
({\sl Darboux theorem}) \beq \omega_{can}=\sum_{a=1}^{N}dp_a\wedge
dq^a, \label{7}\eeq
where $(p_a, q^a)$ are the local coordinates of the symplectic manifold.\\

{\sl The vector field ${\bf V}$ defines a symmetry of the Hamiltonian system,
 if it preserves both the
Hamiltonian ${\cal H}$ and the symplectic form $\omega$}:
 ${\cal L}_{{\bf V}\omega}=0$, ${\bf V}{\cal H}=0$.
 Hence,
 \beq
 {\bf V}=\{{\cal J},\;\},\quad \{{\cal J},{\cal H}\}=0.
\label{8} \eeq
The $2N$-dimensional  Hamiltonian system is called an {\sl integrable system},
when it  has $N$ functionally independent constants of motion
being in involution ({\sl Liouville theorem}),
\beq
\{{\cal J}_a,{\cal J}_b\}=0,\quad \{{\cal H},{\cal J}_b\}=0,\quad
{\cal H}={\cal J}_1,
\quad a,b=1,\ldots, N.
\label{9}\eeq

When the constants of motion are noncommutative, the integrability of the system
needs more than $N$ constants of motion.
If
\beq
\{{\cal J}_{\mu } , {\cal J}_{\nu } \}=
f_{\mu\nu }({\cal J}),\quad {\rm corank }\; f_{\mu\nu}=K_0,
\quad \mu,\nu=1,\ldots , K\geq K_0\;,
\label{10}\eeq
then the system is integrable, if  $2N=K+K_0$ .
The system with $K+K_0\geq 2N$ constants of motion is sometimes called a
{\sl superintegrable system}.
\\

The cotangent bundle  $T^*M_0$ of any manifold  $M_0$ (parameterized by
 local coordinates $q^i$) could be equipped with the canonical
symplectic structure (\ref{7}).

The dynamics of a free particle moving on $M_0$ is given by the Hamiltonian system
\beq
\left( T^*M_0,\;\; \omega_{can},\;\; {\cal H}_0=\frac12 g^{ab}(q)p_a p_b\right),
\label{11}\eeq
where  $g^{ab}g_{bc}=\delta^a_c$,  and  $g_{ab}dq^adq^b$ is
 a metric on $M_0$.\\

The interaction with a potential field could be incorporated in this system
by the appropriate change of Hamiltonian,
\beq
{\cal H}_0\quad\to \quad {\cal H}=\frac12 g^{ab}(q)p_a p_b + U(q),
\label{12}\eeq
where $U(q)$ is a scalar function called potential.
 Hence, the corresponding Hamiltonian system is
given by the triplet $( T^*M_0,\;\; \omega_{can},\;\; {\cal H})$.\\

In contrast to the potential field,
the interaction with a magnetic field requires a change of symplectic structure.
Instead of the canonical symplectic structure $\omega_{can}$, we have to choose
\beq
\omega_{F}=\omega_{can} +{ F},\qquad  F=\frac 12 F_{ab}(q)dq^a\wedge dq^b,\; dF=0
\label{13}\eeq
where $F_{ab}$ are components of  the magnetic field strength.

Hence, the resulting system is given by the triplet
$(T^*M_0,\;\; \omega_{F},\;\; {\cal H})$.
Indeed, taking into account that the two-form $F$  is locally exact, $F=dA$,
 $A=A_a(q)dq^a$,
we could pass to the canonical
coordinates $(\pi_a=p_a+A_a,\; q^a )$. In these coordinates the Hamiltonian
system assumes  the conventional form
$$ \left(T^*M_0,\;\; \omega_{can}=d\pi_a\wedge dq^a ,\;
\; {\cal H}=\frac 12 g^{ab}(\pi_a-A_a)(\pi_b-A_b)+U(q)\right).$$
Let us also remind, that in the three-dimensional case the magnetic field
could be identified with vector, whereas
in the two-dimensional case it could be identified with (pseudo)scalar.\\

The generic Hamiltonian system could be described by the following (phase space) action
\beq
{\cal S}=\int dt\left({\cal A}_i(x)\dot x^i-{\cal H}(x)\right),
\label{14}\eeq
where ${\cal A}={\cal A}_idx^i$ is a symplectic one-form: $d{\cal A}=\omega$.
Indeed, varying the action, we get the equations
\beq
\delta {\cal S}=0,\qquad \Leftrightarrow
\qquad {\dot x}^i\omega_{ij}(x)=\frac{\partial H}{\partial x^i},
\quad
\omega_{ij}=\frac{\partial A_i}{\partial x^j}- \frac{\partial A_j}{\partial x^i}.
\nonumber\eeq
Though  ${\cal A}$ is defined up to closed (locally exact) one-form,
${\cal A}\to {\cal A}+df(x)$, this arbitrariness has no   impact
in the equations of motion. It change the
Lagrangian on the total derivative $f_{,i}{\dot x}^i=df(x)/dt$.

As an example, let us consider the particle in a magnetic field.
The symplectic one-form corresponding to the symplectic structure
(\ref{13}), could be chosen in the form
$
{\cal A}=(p_a+A_a)dq^a$, $d{\cal A}=\omega_F$.
Hence, the action (\ref{14}) reads
\beq
{\cal S}=\int dt\left((p_a+A_a){\dot q}^a - \frac12 g^{ab}(q)p_a p_b- U(q) \right).
\label{16}\eeq
Varying this action by $p$, we get, on the extrema, the
conventional second-order action for the
system in a magnetic field
\beq
{\cal S}_0=\int dt\left(\frac 12 g_{ab}{\dot q}^a{\dot q}^b
+A_a{\dot q}^a - U(q) \right).
\label{17}\eeq
The presented manipulations are nothing but the Legendre transformation from the
Hamiltonian formalism  to the Lagrangian one.

\subsection*{Particle in the Dirac monopole field}
Let us consider the special case of a system on three-dimensional
space moving in the  magnetic field of a Dirac monopole. Its
symplectic structure is given by the expression \beq
\omega_{D}=dp_a\wedge dq^a+s\frac{q^a}{2|q|^3}\epsilon_{abc}
dq^b\wedge dq^c. \label{18}\eeq
The corresponding Poisson brackets
are given by the relations \beq \{p_a,q^b\}=\delta^{b}_{a},\quad
\{q^a, q^b\}=0,\quad\{p_a, p_b\}
=s\epsilon_{abc}\frac{q^c}{|q|^3}. \label{19}\eeq It is clear that
the monopole field does not break the rotational invariance of the
system. The vector fields generating  $SO(3)$ rotations are given
by the expressions
\beq {\bf
V}_a=\epsilon_{abc}q^b\frac{\partial}{\partial q^c}-
\epsilon_{abc}p_b\frac{\partial}{\partial p_c},\quad [{\bf
V}_a,{\bf V}_b]= \epsilon_{abc}{\bf V}_c. \label{20}\eeq The
corresponding Hamiltonian generators could be easily found as well
\beq \imath_{{\bf V}_a}\omega_D=d{\cal J}_a,\qquad \{{\cal
J}_a,{\cal J}_b\}= \epsilon_{abc}{\cal J}_c \;, \nonumber \eeq
where
\beq {\cal J}_{a}=\epsilon_{abc}q^b
p_c+s\frac{q^a}{|q|},\quad J_a q^a=s|q|. \label{21}\eeq
 Now, let
us consider the system given   by the symplectic structure
(\ref{18}), and by the $so(3)$-invariant Hamiltonian \beq {\cal
H}=\frac{p_ap_a}{2g}+U(|q|)\;,\quad \{{\cal J}_a,{\cal H}\}=0\;,
\label{23}\eeq where $g(|q|)dq^a dq^a$ is $so(3)$-invariant metric
on $M_0$. In order to find the trajectories of the system, it is
convenient
 to  direct
 the $q^3$ axis  along the vector ${\bf J}=({\cal J }_1, {\cal J }_2,{\cal J }_3)$,
i.e. to assume that $ {\bf J}=J_3\equiv J$. Upon this choice of the coordinate system
one has
 \beq
\frac{ q^3}{ |q|}=\frac sJ.
\label{tan}\eeq
 Then, we introduce the angle
\beq
 \phi={\rm arctan} \frac{ q^1}{ q^2}\;,\quad
  \frac{d\phi}{dt}=\frac{J^2-s^2}{Jg |q|^2},\label{teq}
\eeq
and get, after obvious
manipulations
 \beq {\cal E}=\frac{J^2-s^2}{2g|q|^2}
 +\frac{(J^2-s^2)^2}{2Jg|q|^2}\left(\frac{d{|q|}}{d\phi}\right)^2
 +U(|q|).\label{teq}
\eeq
Here ${\cal E}$ denotes the energy of the system.\\
From the expression  (\ref{teq}) we find,
\beq
{\phi}=\left({J-\frac{s^2}{J}}\right)\int\frac{d{ |q|}}{\sqrt{2g|q|^2({\cal
E}-U)-J^2+s^2}}.
\label{tg}\eeq
It is seen that, upon the replacement
\beq
U(q)\to U(q)+\frac{s^2}{2g|q|^2},
\label{modif}\eeq
we shall eliminate in (\ref{tg})  the dependence on $s$,
i.e. on a monopole field.
The only impact of the monopole field on the trajectory will be the
shift of the orbital
plane given by (\ref{tan}).

Let us summarize our considerations. Let us consider the $so(3)$-invariant three-dimensional
system
\beq
\omega_{can}=d{\bf p}\wedge d{\bf q},\quad
{\cal H}=\frac{{\bf p}^2}{2g}+U(|{\bf q}|)\;,
\quad \{{\bf J}_0,{\cal H}\}=0\;,\;{\bf J}_0={\bf p}\times {\bf q}.
\label{24}\eeq
Then, replacing it by the following one:
\beq
\omega_{can}+
s\frac{{\bf q}\cdot( d{\bf q}\times d{\bf q})}{2|{\bf q}|^3}
\;,
\quad {\cal H}=\frac{1}{2g}\left( {\bf p}^2+ \frac{s^2}{|{\bf q}|^2}\right)+
U(|{\bf q}|)\;,\;{\bf J}={\bf J}_0 +s\frac{{\bf q}}{|{\bf q}|},
\label{25}\eeq
we shall preserve the form of the  orbit of the initial system, but shift it along
 ${\bf J}$ in accordance with (\ref{tan}).

One can expect that, when the initial system has a symmetry, additional with respect
to the rotational one,
the latter system  will also inherit it.
For  the Coulomb system, $U=\gamma/|{\bf q}|$, this is indeed a case.
The modified system (which is  known as a MIC-Kepler system)
 possesses the hidden symmetry given by the analog of the Runge-Lenz
 vector, which is completely similar to the Runge-Lenz vector of the
 Kepler system \cite{Z}.

\subsection*{K\"ahler manifolds}
 One of the most important classes of symplectic manifolds is that of
K\"ahler manifolds. The  Hermitean manifold
 $(M, g_{a\bar b}dz^adz^b)$ is called K\"ahler manifold, if
  the imaginary part of the Hermitean structure is
a symplectic two-form (see, e.g. \cite{kobnom,perelomov}):
\beq
\omega=ig_{a\bar b}dz^a\wedge d{\bar z}^b \; :
\;\; d\omega=0\;, \det\; g_{a\bar b}\neq 0.
\label{26}\eeq
The Poisson brackets associated with this symplectic structure read
\begin{equation}
   \{ f,g\}_0 =
 i\frac{\partial  f}{\partial {\bar z}^a}g^{{\bar a}b}
\frac{\partial  g}{\partial z^b} -
i\frac{\partial g}{\partial z^b}g^{{\bar a}b}
\frac{\partial  f }{\partial {\bar z}^a},\quad{\rm where }\quad
g^{{\bar a}b}
g_{b{\bar c}}=\delta^{\bar a}_{\bar c}.      \label{p0}
\end{equation}
>From the closeness of (\ref{26}) it  immediately follows,
 that the K\"ahler metric can be
 locally represented in the form
  \begin{equation}
  g_{a \bar b}dz^ad{\bar z}^b =
\frac{\partial^2 K}{\partial z^a \partial{\bar z}^{b}}dz^ad{\bar z}^b,
 \label{27} \end{equation}
where $K(z, \bar z)$ is some real function called the K\"ahler potential.
The K\"ahler potential  is defined modulo  holomorphic
and antiholomorphic functions
\beq
K(z,\bar z)\to K(z,\bar z) + U(z) +{\bar U}(\bar z).
\label{28}\eeq
The local expressions
for the differential-geometric objects on K\"ahler manifolds are also very
simple.
For example, the  non-zero components of the metric connections (Cristoffel symbols)
look as follows:
\beq
\Gamma^a_{bc}=g^{\bar n a}g_{b{\bar n}, c},\quad \Gamma^{\bar a}_{\bar b\bar c}=
{\overline\Gamma^a_{bc}},
\label{29}\eeq
while the non-zero components of the curvature tensor read
\beq
R^a_{bc\bar d}=-(\Gamma^a_{bc})_{,d},\quad
R^{\bar a}_{\bar b\bar c d}= {\overline R^a_{bc\bar d}}.
\label{30}\eeq
The isometries
of K\"ahler manifolds are
 given by the {\it holomorphic  Hamiltonian vector fields}
 \begin{equation}
{\bf V}_{\mu}=
    V_\mu^{a}(z)\frac
{\partial}{\partial z^a}+
{\bar V}_\mu^{\bar a}(\bar z) \frac{\partial}
{\partial \bar z^a},
\; , \;{\bf V}_\mu=\{\ch_\mu, \}_0 ,
\label{33}\end{equation}
where $\ch_\mu$  is a real function, $\ch_\mu=\bar\ch_\mu$, called Killing potential.
One has
$$
   [{\bf V}_{\mu},{\bf V}_{\nu}]=
C_{\mu \nu}^{\lambda}{\bf V}_{\lambda},\quad  \{ \ch_{\mu},\ch_{\nu}\}_0=
C_{\mu \nu}^{\lambda}
\ch_{\lambda}+{\rm const}\;,
$$
and
$$
\frac{\partial^2 \ch_\mu}{\partial z^a \partial z^b} -
\Gamma^c_{ab}\frac{\partial \ch_\mu}{\partial z^c}=0.
$$
The dynamics of a particle moving on the K\"ahler manifold in the presence of a constant
magnetic field is described
by  the  Hamiltonian system
\begin{equation}
\Omega_B=
dz^a\wedge d\pi_a +
d{\bar z}^{a}\wedge d{\bar\pi}_{a} +iBg_{a\bar b}dz^a\wedge
d{\bar z}^b,\quad  {\cal H}_0= g^{a \bar b}\pi_a{\bar \pi}_b
\label{ssB}\end{equation}
The isometries of a K\"ahler
 structure define the
Noether constants of motion
 \begin{equation}
{\cal J}_{\mu}\equiv J_\mu+ B\ch_\mu=V_\mu^{a}\pi_a +
 {\bar V}_{\mu}^{\bar a} {\bar\pi}_{\bar a} +B\ch_\mu :
\left\{
\begin{array}{c}
  \{{\cal H}_0, J_{\mu}\}=0, \\
  \{J_\mu, J_\nu\}=C_{\mu\nu}^\lambda J_\lambda .
\end{array}\right.\label{jmu}\end{equation}
One can easily check that the vector fields
 generated by ${\cal J}_\mu$ are independent of $B$
\begin{equation}
{\bf{ V}}=V^a(z)\frac{\partial}{\partial z^a}-V^a_{,b}\pi_a
\frac{\partial}{\partial \pi_a}+
{\bar V}^a(\bar z)\frac{\partial}{\partial\bar z^a}
-{\bar V}^a_{,\bar b}\bar\pi_a
\frac{\partial}{\partial \bar\pi_a}\quad.
\label{34}\end{equation}
Hence, the inclusion of a
 constant magnetic field preserves
the whole symmetry algebra of a free
particle moving on a K\"ahler manifold.

\subsection*{Complex projective space}
The most known nontrivial example of a K\"ahler manifold is the
 complex projective space  $\DC P^N$.
 It is  defined as a space of complex lines in
 $\DC^{N+1}$: $u^{\tilde a}\sim \lambda u^{\tilde a}$, where $u^{\tilde a}$,
 ${\tilde a}=0,1,\ldots, N$ are the Euclidean coordinates of $\DC^{N+1}$, and
 $\lambda\in \DC-\{0\}$. Equivalently, the complex projective space is the coset space
  $\DC P^N=SU(N+1)/U(N)$.

The complex projective space $\DC P^N$ could be
 covered by $N+1$ charts marked by the
indices ${\tilde a}=0,a$.
The zero chart could be parameterized by the functions (coordinates)
 $z^a_{(0)}=u^a/u^0$,
 $a=1,\ldots N$; the  first chart  by $z^a_{(1)}=z^a/z^1$, $a=0,2,3\ldots, N$,
  and so on.

Hence, the transition function
from the ${\tilde b}$-th chart to
the ${\tilde c}$-th one has the form
\beq
z^{\tilde a}_{({\tilde c})}=
\frac{z^{\tilde a}_{({\tilde b})}}{z^{\tilde c}_{({\tilde b})}},
\quad {\rm where}\quad z^{\tilde a}_{({\tilde a})}=1 .
\label{trans}\eeq
One can  equip the $\DC P^N$ by the K\"ahler metric, which is
known under the name of
Fubini-Study metric  \begin{equation}
g_{a\bar b}dz^a dz^b=\frac{dzd{\bar z}}{1+z\bar z}-
 \frac{({\bar z}dz)(zd{\bar z})}{(1+z\bar z)}.
 \label{35} \end{equation}
  Its  K\"ahler potential is given by the expression
  \beq
  K=\log (1+ z\bar z).
 \label{36} \eeq
 Indeed, it is seen that upon transformation from  one chart
to the other, given by (\ref{trans}), this  potential changes
by holomorphic and anti-holomorphic functions,
i.e. the Fubini-Study metric is globally defined on $\DC P^N$.

The Poisson brackets on $\DC P^N$ are defined by the following relations:
\beq
\{z^a,\bar z^b\}=(1+z\bar z)(\delta^{a\bar b}+z^a{\bar z}^b),\qquad
\{z^a,z^b\}=\{\bar z^a,\bar z^b\}=0.
\label{37}\eeq
It is easy to see that  $\DC P^N$ is a constant curvature space, with
the symmetry algebra  $su(N+1)$.
This algebra is  defined  by the Killing  potentials
\begin{equation}
\ch_{\bar a b}=\frac{z^a \bar z^b- N \delta_{\bar a b}}{1+ z\bar z},\quad
\ch^-_a=\frac{z^a}{1+z\bar z},\quad
\ch^+_a=\frac{\bar z^a}{1+ z\bar z}.
\label{38}\end{equation}
The manifold  $\DC P^1$ (complex projective plane)
is isomorphic to the two-dimensional sphere $S^2$.
Indeed, it is covered by the two charts, with the transition function
$z\to 1/z$.
The symmetry algebra of $\DC P^1$ is $su(2)= so(3)$
\beq
\{x^i,x^j\}=\epsilon^{ijk}x^k,\quad i,j,k=1,2,3
\eeq
where   the Killing potentials $x^i$ look as follows:
\beq
x^1+ix^2=\frac{2z}{1+z\bar z},\quad x^3=\frac{1-z\bar z}{1+z\bar z}
.\label{40}\eeq
It is seen that these Killing potentials satisfy the condition
$$
x^ix^i=1,
$$
i.e. $x^i$ defines the sphere $S^2$ in the three-dimensional ambient space $\DR^3$.
It is straightforwardly checked that $z$ are the coordinates of the sphere
in the  stereographic projection on $\DR^2=\DC$. The  real part
of the  Fubini-Study  structure  gives the linear element of $S^2$,
 and the imaginary  part coincides with the volume element of $S^2$.

On the other hand, these expressions give the embedding of the $S^2$ in $S^3$
(with ambient coordinates $u^1, u^2$) defining  the  so-called
first Hopf map  $S^3/S^1=S^2$. Below we shall
 describe this map in more detail.
\subsection*{Hopf maps}
The Hopf maps (or Hopf fibrations) are the fibrations of the sphere over a sphere,
\beq
S^{2p-1}/S^{p-1}=S^{p},\qquad p=1,2,4,8.
\eeq
These fibrations reflect the existence of real ($p=1$), complex ($p=2$),
quaternionic ($p=4$)
and octonionic ($p=8$) numbers.

We are interested in the so-called zero-th, first and second Hopf maps:
\bea
 &S^1/S^0=S^1\quad ({\rm zero}\;{\rm Hopf}\;{\rm map})&\nonumber\\
&S^3/S^1=S^2\quad ({\rm first}\;{\rm Hopf}\;{\rm map})&
  \label{hm0}\\
&S^7/S^3=S^4\quad ({\rm second}\;{\rm Hopf}\;{\rm map})&.\nonumber
 \eea
Let us describe the Hopf maps in explicit terms.
For this purpose, we consider the  functions ${\bf x}(u,\bar u), x_0(u,\bar u )$
\beq
{\bf x}=2u_1{\bar u}_2,\quad x_{p+1}=u_1{\bar u}_1-{u_2\bar u}_2,
\label{hm}\eeq
where $u_1,u_2$, could be real, complex or  quaternionic numbers.
So, one can consider them as a coordinates of
the  $2p$-dimensional  space $\DR^{2p}$,
 where $p=1$ when $u_{1,2}$ are real numbers;
$p=2$ when $u_{1,2}$ are complex numbers;
$p=4$ when $u_{1,2}$ are quaternionic numbers;
$p=8$ when $u_{1,2}$ are octonionic ones.

In all  cases  $x_{p+1}$ is a real number, while ${\bf x}$ is,
respectively,
a real number ($p=1$), complex number ($p=2$), quaternion($p=4$),
   or octonion ($p=8$).
 Hence, $(x_0, {\bf x})$ parameterize the $(p+1)$-dimensional space $\DR^{p+1}$.

 The functions  ${\bf x}, x_{p+1}$ remain invariant under transformations
\beq
u_a\to {\bf g} u_a,\quad {\rm where}\quad g{\bar g}=1.
\label{bundle}\eeq
Hence
\bea
&{g}=\pm 1\quad {\rm for}\quad p=1&\label{b0}\\
&g=\lambda_1+i\lambda_2\;\;,\;\;
\lambda_1^2+\lambda_2^2=1\quad{\rm for}\quad p=2 &\label{b1}\\
&g=\lambda_1+i\lambda_2+j\lambda_3+k\lambda_4\;\;,\;\;
\lambda_1^2+\ldots + \lambda_4^2=1\quad{\rm for}\quad p=4 .&\label{b2}
\eea
and similarly for the octonionic case $p=8$.\\
So, $g$ parameterizes the spheres $S^{p-1}$ of unit radius.
Notice that $S^1, S^3, S^7$ are the only parallelizable  spheres.
We shall also use the following isomorphisms
 between these spheres and groups: $S^0=Z_2$, $S^1=U(1)$, $S^3=SU(2)$.

We get that (\ref{hm}) defines the fibrations \beq
\DR^2/S^0=\DR^2,\quad\DR^4/S^1=\DR^3,\quad\DR^8/S^3=\DR^5, \quad
\DR^{16}/S^7=\DR^9. \eeq
One could  immediately check that the
following equation holds:
\beq {\bf x}{\bf \bar x}+x^2_{p+1}=(u_1
{\bar u}_{1}+u_2{\bar u}_2)^2. \label{hm1}\eeq Thus, defining the
$(2p-1)$- dimensional sphere in $\DR^{2p}$ of the radius $r_0$:
$u_a\bar u_a =r_0$, we will get the $p$-dimensional sphere in
$\DR^{p+1}$ with radius $R_0=r_0^2$ \beq u_1 {\bar u}_{1}+u_2{\bar
u}_2=r^2_0\quad \Rightarrow\quad {\bf x}{\bf \bar x}+x^2_0=
r^4_0\; . \label{hm2}\eeq So, we arrive at the Hopf maps given by
(\ref{hm0}). The last, fourth Hopf map, $S^{15}/S^7=S^8$,
corresponding to $p=8$, is  related to octonions in the same
manner.

For our purposes it is convenient to describe the the expressions (\ref{hm})
 in a less unified
way.
For the  zero Hopf map it is
convenient to consider the initial and resulting  ambient spaces
$\DR^2$ as complex spaces $\DC$, parameterized by the single complex coordinates
$w$ and $z$.
In this case the map (\ref{hm}) could be represented
 in the form
\beq
w=z^2,
\label{bohlin0}\eeq
which is known as a Bohlin (or Levi-Civita) transformation
relating the Kepler problem  with the circular oscillator.

For the  first and second Hopf maps it is convenient to represent the
transformation
(\ref{hm}) in the following  form:
\beq
{\bf x}=u\bg{\bar u}.
\label{ks0}\eeq
Here, for the  first Hopf map  ${\bf x}=(x^1,x^2, x^3)$
 parameterizes $\DR^3$, and $u_1, u_2$ parameterize
$\DC^2$, and $\bg =(\sigma^1,\sigma^2,\sigma^3)$ are Pauli matrices.
This transformation is also
known under the name of Kustaanheimo-Stiefel transformation.
For the second Hopf map
${\bf x}=(x^1,\ldots, x^5)$ parameterizes $\DR^5$, and $u_1,\ldots, u_4$
 parameterize
$\DC^4=\DH^2$, and $\bg =(\gamma^1,\ldots,\gamma^4,
\gamma^5=\gamma^1\gamma^2\gamma^3\gamma^4)$,
where $\gamma^1,\ldots, \gamma^4$ are Euclidean four-dimensional gamma-matrices.
The latter transformation is sometimes called Hurwitz transformation,
 or ``generalized Kustaanheimo-Stiefel" transformation.

\section{Hamiltonian reduction}
A Hamiltonian system which has a constant(s) of motion, can be  reduced
to a lower-dimensional one. The corresponding procedure is called Hamiltonian reduction.
Let us explain the meaning of this procedure
in the simplest case
 of the Hamiltonian
reduction by a single constant of motion.

Let $(\omega, {\cal H})$ be a given $2N$-dimensional
 Hamiltonian  system, with the phase space (local) coordinates $x^A$,
  and let ${\cal J}$  be its
constant of motion, $\{{\cal H}, {\cal J}\}=0$.
We go from the local coordinates $ x^A$ to
another set
 of coordinates, $({\cal H}, y^i, u )$, where
 $y^i=y^i(x)$ are $2N-2$ independent  functions, which commute with
${\cal J}$,
\beq
\{y^i, {\cal J}\}=0,\qquad i=1,\ldots, 2N-2.
\label{41}\eeq
In this case
the latter coordinate, $u=u(x)$,
 necessarily has a non-zero Poisson bracket with
${\cal J}$ (because the Poisson brackets are nondegenerate):
\beq
\{u(x),{\cal J}\}\neq 0.
\label{42}\eeq
Then, we immediately get that in these coordinates
the Hamiltonian is independent of $u$
\beq
\{{\cal J}({\cal H},y, u) ,{\cal H}\}=
\frac{\partial{{\cal H}}}{\partial u}\cdot \{u,{\cal J}\}\neq 0,
\;\Rightarrow \;{\cal H}=
{\cal H}({\cal J}, y).
\label{43}\eeq
On the other hand, from the Jacobi identity we get
\beq
\{\{y^i, y^j\},{\cal J}\}=
\frac{\partial \{y^i, y^j\}}{\partial u}\{u,{\cal J}\}=0\;\Rightarrow \;
\{y^i, y^j\}=\omega^{ij}(y, {\cal J}).
\label{44}\eeq
Since ${\cal J}$ is a constant of motion, we can fix its value
\beq
{\cal J}=c,
\label{45}\eeq
and describe the system in terms of the local coordinates $y^i$ only
\beq
\left(\omega (x), {\cal H}(x)\right)\;\rightarrow
\left(\omega_{red}(y, c)=\omega_{ij}(y,c)dy^i\wedge dy^j,
 {\cal H}_{red}={\cal H}(y, c) \right).
\label{46}\eeq
Hence, we reduced the initial $2N$-dimensional Hamiltonian system
 to a $(2N-2)$-dimensional one.

Geometrically, the Hamiltonian reduction by ${\cal J}$ means that
we fix
 the $(2N-1)$- dimensional level surface $M_c$ by the Eq.(\ref{45}),
 and then  factorize it
  by the action of a vector field $\{{\cal J},\;\}$,
  which is tangent  to $M_c$.
 The resulting space ${\cal M}_0=M_c/\{{\cal J},\;\}$ is a phase space of the
 reduced system.

The Hamiltonian reduction by the $K$ commuting constants of motion ${\cal J}$,
$\{{\cal J}_\alpha ,{\cal J}_\beta\}=0$ is completely similar to the above procedure.
It reduces the $2N$ dimensional Hamiltonian system to a $2(N-K)$
dimensional one.\\

When the constants of motion do not commute with each other, the
reduction procedure is a bit more  complicated.

 Let the initial Hamiltonian system have $K$ constants of motion,
 \beq
 \{{\cal J}_{\alpha}, {\cal H}\}=0,\quad
\{{\cal J}_{\alpha}, {\cal J}_{\beta}\}=\omega_{\alpha\beta}({\cal J}), \qquad
{\rm corank}\;\omega_{\alpha\beta}\vert_{{\cal J}_\alpha=c_\alpha}=K_0.
 \label{47}\eeq
 Hence,  one could choose the $K_0$ functions, which commute
  with the whole set of the constants of motion
 \beq
 \widetilde{\cal J}_{\tilde\alpha}({\cal J})\; :\;
 \{\widetilde{\cal J}_{\tilde\alpha},{\cal J}_\beta\}
 \vert_{{\cal J}=c}=0,\qquad {\tilde\alpha}=1,\ldots K_0.
 \eeq
The vector fields
 $\{\widetilde{\cal J}_{\tilde\alpha},\;\}$ are tangent to the level surface
 \beq
 M_c\;: {\cal J}_\alpha=c_\alpha\,\qquad {\rm dim} M_c=2N-K.
 \eeq
Factorizing $M_c$ by the action of the commuting vector fields
$\{\widetilde{\cal J}_{\tilde\alpha},\;\}$, we arrive at the phase space
of the reduced system, ${\cal M}_0=M_c/\{{\cal J},\;\}$,
whose dimension is given by the expression
\beq
{\rm dim M}_0=2N-K-K_0\;.
\eeq
In contrast to  the commuting case, the reduced system could
depend on the parameters ${\widetilde c}_{\tilde\alpha}$ only.

Notice that the  Hamiltonian system could also possess a discrete symmetry.
In this case the reduced system has the same dimension as the previous one.
To be more precise, the reduction by the discrete symmetry group
could be described by a {\sl local} canonical transformation.
However, the quantum mechanical counterpart of this canonical transformation
could yield a system with non-trivial physical properties.

Below, we shall illustrate the procedure of (Hamiltonian) reduction
by discrete, commutative, and noncommutative symmetry generators
on examples related to Hopf maps.

\subsection*{Zero Hopf map. Magnetic flux tube}
The transformation of the Hamiltonian system associated with the
zero Hopf map  corresponds to the reduction of the system by the discrete
group $Z_2$. It is a (local) canonical transformation.
As a consequence, the resulting system has the same dimension
as the initial one.

Let us consider the  Hamiltonian system with four-dimensional
phase space, parameterized by the pair of  canonically
conjugated complex coordinates,
$(\omega=d\pi\wedge dz+d\bar\pi\wedge d\bar z, {\cal H} )$,
which is invariant under the following action of  $Z_2$ group:
$$
{\cal H}(z,\bar z, \pi,\bar\pi)={\cal H}(-z,-\bar z, -\pi,-\bar\pi),\quad
\omega(\pi,\bar\pi, z, \bar z)=\omega(-\pi,-\bar\pi,- z, -\bar z).$$
We can pass now to the coordinates, which are invariant
under this transformation
(clearly, it is associated with the zero Hopf map)
\bea
&w=z^2,\quad p={\pi}/{2z}&\label{b}\\
&\omega=
d\pi\wedge dz+d\bar\pi\wedge d\bar z=dp\wedge dw+d\bar p\wedge d\bar w \; .&
\eea
However, one can see that the angular momentum of
the initial systems looks as a doubled
 angular momentum of
the transformed one
 \begin{equation} J= i(z\pi -\bar z\bar\pi )=2i(w p-\bar w\bar p).
     \label{j0}
 \end{equation}
This indicates that the global properties of these two systems could
 be essentially  different.
 This difference has to be reflected in
  the respective quantum-mechanical systems.\\

Let us consider the Schr\"odinger equation
\begin{equation}
{\cal H}(\pi,\bar\pi, z,\bar z)\Psi(z,\bar z)=E\Psi(z,\bar z),
\qquad \pi=-i\partial_z,\;\bar\pi=-i\partial_{\bar z},
\label{1b}\end{equation}
with the wavefunction which obeys the condition
\beq
  \;\Psi(|z|, {\rm arg}\; z +2\pi)=\Psi(|z|, {\rm arg}\; z ).
\eeq
Let us reduce it  by the action of $Z_2$ group,
 restricting ourselves to even  ($\sigma=0$)
or odd ($\sigma=\frac 12$)  solutions  of Eq. (\ref{1b})
 \begin{equation}
  \Psi_\sigma (z, \bar z) =
\psi_\sigma (z^2, {\bar z}^2){\rm e}^{2i\sigma {\rm arg}\;z},\quad \sigma=0,1/2,
\label{3b}\end{equation}
and then perform the Bohlin transformation
(\ref{b}).
According to Eq.(\ref{3b}), the wave functions  $\psi_\sigma$
satisfy the condition
\beq
\psi_{\sigma}(|w|, {\rm arg} w + 2\pi)=
\psi_{\sigma}(|w|, {\rm arg} w ),\eeq
 which implies that the range  of
definition ${\rm arg}\;w\in[0,4\pi)$ can be restricted,
without loss of generality, to ${\rm arg} w\in[0,2\pi)$.
In terms of $\psi_\sigma$ the Schr\"odinger equation (\ref{1b})
reads
\begin{equation}
{\cal H}({\hat p}_\sigma ,{\hat p}^{+}_\sigma ,
 w,\bar w)\psi_\sigma(w,\bar w)=E \psi_\sigma , \qquad
 {\hat p}_\sigma = -i\partial_w-\frac{i\sigma}{w}.\label{4b}
\end{equation}

Equation (\ref{4b}) can be interpreted as the Schr\"odinger equation of a
particle with electric charge  $e$ in the static
magnetic field given  by  the potential
$ A_w=\frac{i\sigma}{e w}$, $\sigma =0,{1}/{2}$.
It is a potential of an
infinitely thin solenoid- ``magnetic flux tube" (or magnetic vortex,
in the two-dimensional interpretation): it has  zero strength
of the magnetic field $B=rot A_w=0$ ($w\in\dot{\DC}$)
 and nonzero magnetic flux $2\pi\sigma/e$.

In accordance with (\ref{j0}), the angular momentum transforms as follows:
\begin{eqnarray} &J\to &2J_\sigma
,\;\;J_\sigma=\frac{i}{\hbar}\left( w{\hat p}_{\sigma}- {\bar w}{\hat
p}^+_{\sigma}\right), \label{j}
\end{eqnarray}
 where  ${J}_\sigma$  is the angular
momentum operator of the reduced system.
Hence, the  eigenvalues  of the angular momenta
  of the reduced and initial systems,
  $m_\sigma$  and $M$, are related   by the expression $M= 2m_\sigma$,
from which  it follows that
 \begin{equation} m_\sigma= \pm\sigma, \pm (1+\sigma),
\pm (2 +\sigma),\ldots .
 \end{equation}
 Hence, the $Z_2$-reduction related to zero Hopf map transforms the
  even states of the initial system to the complete basis of the resulting one.
  The odd states of the initial system yield the wave functions
  of the resulting system in the presence of magnetic flux generating spin
   $ 1/2$.
Similarly to the above consideration,  one can show that the reduction of
the two-dimensional system by the $Z_N$ group yields the $N$ systems
with the fractional spin $\sigma=0, 1/N, 2/N , \ldots, (N-1)/N$
(see \cite{ntt}).

\subsection*{$1$st Hopf map. Dirac monopole}
Now we consider the Hamiltonian  reduction by the action of the
$U(1)$ group, which is
associated with the first Hopf map. It is known
under the name of Kustaanheimo-Stiefel transformation.

Let us consider the Hamiltonian system  on
the four-dimensional Hermitean space
$(M_0,\;g_{a\bar b}dz^a d{\bar z}^b)$, ${\rm dim}_{\DC}M_0=2$,
\begin{equation}
T^*M_0,\quad \omega=
dz^a\wedge d\pi_a +
d{\bar z}^{a}\wedge d{\bar\pi}_{a},\quad {\cal H}=  g^{a \bar b}\pi_a{\bar \pi}_b
+ V(z, {\bar z}).
\label{48}\eeq
We define, on the $T^*M_0$ space, the Hamiltonian action of the $U(2)$ group
given by the generators
\bea
&{\bf J}=iz{\bs}\pi-i\bar\pi{\bs}\bar z,
\quad J_0=iz\pi-i\bar z\bar\pi\; :&\\
&\{J_0, {J}_k\}=0,\quad \{J_k,J_l \}=2\epsilon_{klm}J_m,&
\eea
where $\bs$ are  Pauli matrices.\\
Let us consider the Hamiltonian reduction of the phase space $(T^*M_0, \omega)$
by the (Hamiltonian) action of the $U(1)=S^1$ group given
by the generator $J_0$.
Since $J_0$ commutes with $J_i$,
 the latter will generate the Hamiltonian
 action of the $su(2)=so(3)$ algebra on the reduced space as well.

In order to perform the Hamiltonian reduction, we have to fix the level
surface
 \beq
J_0=2s,\label{49} \eeq
and then factorize it  by  the
action of the vector field $\{J_0,\;\}$.

 The resulting six-dimensional
phase space $T^*M^{\rm red}$ could  be parameterized by the following
$U(1)$-invariant
 functions:
\beq
 {\bf y}=z\bs{\bar z},\quad
 {\bp}=\frac{z\bs\pi +\bar\pi\bs\bar z}{2z\bar z}\;
:\quad \{{\bf y}, J_0\}=\{{\bp}, J_0\}=0.
\label{ksc}\end{equation}
In these coordinates
 the reduced  symplectic structure and the generators of the
 angular momentum
 are given by the expressions (compare with (\ref{18}),(\ref{21}))
\begin{equation}
\Omega_{\rm red}=d{\bp}\wedge d{\bf y} +
s\frac{{\bf y}\cdot (
 d{\bf y }\times d{\bf y})}{2|{\bf y}|^3},
\quad {\bf J}_{red}={\bf J}/2= {\bp}\times{\bf y} + s\frac{{\bf
y}}{ |{\bf y}|  }. \nonumber\end{equation}
Hence, we get the phase space of the Hamiltonian system describing
the motion of a nonrelativistic scalar
particle in the magnetic field of the Dirac monopole.

Let $M_0$ be a $U(2)$-invariant K\"ahler space with a  metric generated by
the K\"ahler potential $K(z\bar z)$ \cite{ny}
 \beq
  g_{a\bar b}=\frac{\partial^2
K(z\bar z)}{\partial z^a{\partial\bar z}^b}= a(z\bar z)\delta_{a\bar b}+
a'(z\bar z){\bar z}^a  z^b,
\eeq
where
\beq a(y)=\frac{dK(y)}{dy}, \quad
a'(y)=\frac{d^2K(y)}{dy^2} .
\label{gk}\nonumber\eeq
Let the potential
be also $U(2)$-invariant, $V=V(z\bar z)$, so that $U(2)$
is a symmetry of the Hamiltonian: $\{J_0, {\cal H}\}=\{J_i,{\cal H}\}=0$.

Hence, the Hamiltonian could also be restricted to  the reduced
 six-dimensional phase space.
The reduced Hamiltonian looks as follows:
\begin{equation}
 {\cal H}_{red}=\frac{1}{a}\left[{y}{\bp}^2 -
 b{({\bf y} {\bp})^2}\right]
+ s^2\frac{1-b y}{a y}+V(y),
\nonumber\eeq
where
\beq
y\equiv |{\bf y}|,\quad
b=\frac{a'(y)}{a+y a'(y)}.
\label{hred}\nonumber\end{equation}

Let us perform the canonical
 transformation $({\bf y},{\bp})\to({\bf{ x}},
{\bf{ p}})$  to the conformal-flat metric
$$
{\bf { x}}= f(y){\bf y},\quad {\bp}= {f} {\bf {  p}}
+\frac{df}{dy} \frac{({\bf y{ p}})}{y}{{\bf y }},
 $$
 where
$$
\left(1+\frac{yf'(y)}{f}\right)^2= 1+\frac{ y a'(y)}{
a}\quad \Rightarrow\quad \left(\frac{d\log x}{dy}\right)^2=
\frac{d\log ya(y)}{ydy},\quad x<1.
$$
 In the new
coordinates the Hamiltonian takes the form
$${\cal H}_{red}=
 \frac{{x}^2(y)}{ya(y)}{\bf{ p}^2}+\frac{s^2}{y(a+y
a'(y))} +V\left(y(x)\right).
$$
In order to express the $y$, $a(y)$, $a'(y)$ via ${ x}$, it is
convenient to introduce the function
$$ {\tilde A}(y)\equiv \int
(a+ya'(y))y f(y)dy
 $$ and consider its Legendre transform ${A}({
x})$,
$$
 { A}({ x})=A(x, y)\vert_{\partial A(x,y)/\partial y},\quad
 A(x,y)={ x} a(y)y-\tilde A(y).\nonumber
$$
 Then, we  immediately get
  \beq \frac{d{ A}({ x})}{d{ x}}=
a(y)y, \quad { x}\frac{d^2{ A}}{d{ x}^2}= {y\sqrt{a(a+y a'(y))}}.
\nonumber\eeq
By the use of  these expressions, we can represent  the reduced
 Hamiltonian  as follows:
 \beq
 {\cal H}_{red}=\frac{{x}^2}{N^2}{\bf{ p}}^2 +
\frac{s^2}{\left(2{x}N'(x)\right)^2} + V\left(y(x)\right),\quad
N^2({ x})\equiv\frac{d{A}}{d{x}}.
\label{rh}\eeq
 The K\"ahler potential
of the initial system is connected with $N$ via the  equations
\beq
\frac{dK}{d{x}}=\frac{N^3(x)}{2{x}^2N'(x)},\quad \frac{d\log
y}{d{x}}=\frac{N}{2{x}^2N'(x)}. \label{corresp}\eeq

Hence, for  $s=0$ we shall get the system (\ref{24}). However, when $s\neq 0$,
by comparing the reduced system with (\ref{25}),
 we conclude that the only K\"ahler space which yields
a ``well-defined system with monopole"  is  flat space.

\subsection*{$\DC^{N+1}\to \DC P^N$ and $T^*\DC^{N+1}\to T^*\DC P^N$}
Now, we  consider the Hamiltonian reduction of the
the  space
 $(\DC^{N+1}, \omega=du^0d\bar u^0+du^a d\bar u^a$),
 to the complex projective space $\DC P^N$.

The  $U(N+1)=U(1)\times SU(N)$ isometries of this space
 are defined by the following
Killing potentials:
\beq
J_0=u\bar u,\quad J_{su(N+1)}=u{\hat T}\bar u,\quad \{J_0,J_{su(N+1)}\}=0,
\nonumber\eeq
where $T=T^{\dagger}$, ${\rm Tr} T=0$ are $(N+1)\times (N+1)$ dimensional traceless
matrices defining the $su(N+1)$ algebra.\\
The Poisson brackets, corresponding to the K\"ahler
structure, are defined by the relations $\{u^0,\bar u^0\}=i$,
$\{u^{},\bar u^{b}\}=i\delta^{ab}$.

Let us perform the Hamiltonian reduction by the action of $J_0$.
The reduced phase space is a $2N$ dimensional one. Let us choose for this space
the following local complex coordinates:
\beq
z^a=\frac{u^a}{u^0}\;:\{z^a, J_0\}=0,\qquad a=1,\ldots, N
\eeq
and fix the level surface
\beq
J_0=r^2_0\;\Rightarrow |u^0|^2=\frac{r^2_0}{1+z\bar z}.
\label{level}\eeq
Then, we immediately
get the Poisson brackets for the reduced space
\beq
\{z^a,\bar z^b\}=\frac{i}{r^2_0}(1+z\bar z)(\delta^{ab}+z^a\bar z^b)\; ,
\qquad\{z^a,z^b\}=\{\bar z^a,\bar z^b\}=0.
\eeq
Hence, the reduced  Poisson bracket are  associated with the K\"ahler structure.
It could be easily seen, that this K\"ahler structure is  given by the
Fubini-Study metric (\ref{35}) multiplied on $r^2_0$.
The restriction of the generators $J_{su(N+1)}$ on the level surface (\ref{level})
yields the expressions (\ref{38}).\\

In the above example $\DC^{N+1}$ and $\DC P^N$ appeared as the phase spaces.
Now, let us show, how to reduce the
 $T^*\DC^{N+1}$ to $T^*\DC P^N$, i.e. let us consider the case when
 $\DC^{N+1}$ and $\DC P^N$ play the role of the configuration spaces
 of the mechanical systems.
Since the dimension of $T^*\DC^{N+1}$ is $4(N+1)$,
 and the dimension of $T^*\DC^N$ is $4N$,
the reduction has to be performed by two commuting generators.

Let us equip the initial space with the canonical symplectic structure (\ref{48}),
and perform the reduction of this phase space by the action of the generators
\beq
J_0=i\pi u-\bar\pi\bar u,\quad h_0=u\bar u\;:\quad\{J_0, h_0\}=0.
\eeq
We choose the following local coordinates of the reduced space:
\beq
z^a=\frac{u^a}{u^0},\quad p_a=g_{a\bar b}(z,\bar z)\left(\frac{\bar\pi^a}{\bar u^0}-
{\bar z}^a\frac{\bar\pi^0}{\bar z^0}\right)\;:\nonumber\eeq
$$\{z^a,J_0\}=\{z^a, h_0\}=
\{ p_a,J_0\}=\{ p_a, h_0\}=0,$$
where $g_{a\bar b}$ is defined by the expression (\ref{35}).
Then, calculating the Poisson brackets between these functions,
and fixing the value of
the generators $J_0, h_0$,
\beq
h_0=r^2_0,\qquad J_0=2s,
\eeq
we get
\beq
\{p_a, z^b\}=\delta^b_a,\quad
 \{p_a,{\bar p}_b\}=i\frac{s}{r^2_0}g_{a\bar b}(z,\bar z).
\eeq
Hence, we arrive at the phase space structure of the particle moving on $\DC P^N$
in the presence of a constant magnetic field with $B_0=s/r^2_0$ strength.

\subsection*{$2$nd Hopf map. $SU(2)$ instanton }
In  the above examples we have shown that the zero Hopf map is related
to the canonical transformation corresponding to the
reduction of the two-dimensional system  by the discrete group $Z_2=S^0$,
and transforms the system with two-dimensional configuration space
to the system of the same dimension, which has a spin $\sigma=0,1/2$.
The first Hopf map  corresponds to the reduction of the
system with four-dimensional configuration space
 by the Hamiltonian action of $U(1)=S^1$ group,
and yields the system moving on the three-dimensional
space in the presence of
 the magnetic field of the
Dirac monopole.
Similarly,  with the second Hopf map one can relate the Hamiltonian
reduction of the cotangent bundle of eight-dimensional space
 (say, $T^*\DC^4= T^* \DH^2$)
 by the action of $SU(2)=S^3$ group.
 When the $SU(2)$ generators
$I_i$ have non-zero values, $I_i=c_i$,$\sum_i |c_i|\neq 0$,
the reduced space is a $(2\cdot 8-3-1=)12$- dimensional one, $T^*\DR^5\times S^2$.
It is the phase space of a coloured particle moving on $\DR^5$ in the presence of
the $SU(2)$ Yang monopole \cite{yang} (here $S^2$ appears as a isospin space).

When $c_1=c_2=c_3=0$, the $J_i$  generators commute with each other,
and
the reduced space is a $(2\cdot 8-2\cdot 3=)10$-dimensional
one, $T^*\DR^5$.
Such a reduction is also known under the name of  Hurwitz transformation
 relating the
eight-dimensional oscillator with the five-dimensional Coulomb problem.

We shall describe a little bit different reduction, associated with the fibration
$\DC P^3/\DC P^1= S^4$ \cite{4hall}. This fibration could be immediately
obtained  by factorization of the second Hopf map
$S^7/S^3=S^4$ by  $U(1)$. Indeed, the second Hopf map is described by the formulae
(\ref{hm}),(\ref{bundle}), where  $S^7$ is embedded
in the two-dimensional quaternionic space
$\DH^2=\DC^4$, parameterized
by four complex (two quaternionic) Euclidean coordinates
\beq { u}_i=v_i+jv_{i+1},\quad i=1,2,\quad {  u}_{1},\;{\bf u}_2\in
\DH, \quad v_1,v_2,v_3,v_4\in \DC\quad . \eeq
Here $S^4$ is embedded in  $\DR^5$ parameterized
by the Eucludean coordinates (${\bf x}, x_5$) given by (\ref{hm}).
This embedding is invariant under
the right action of a $SU(2)$
group  given by (\ref{bundle}),  so that  ${\bf g}$
 defines a three-sphere  (\ref{b2}).
  The complex projective space $\DC P^3$ is defined as $S^7/U(1)$,
   while the inhomogeneous coordinates
$z_a$ appearing in the Fubini-Study metric of $\DC P^3$, are
related to the coordinates of $\DC^4$ as follows:
$z_a=v_a/v_4$, $a=1,2,3$. The expressions
(\ref{hm}) defining $S^4$ are invariant under
$U(1)$-factorization,  while  $S^3/U(1)=S^2$. Thus, we  arrive to
the conclusion that  $\DC P^3$ is the $S^2$-fibration over
$S^4=\DH P^1$.
 The expressions for $z_a$ yield the following
definition of the coordinates of $S^4$: \beq w_1=\frac{\bar z_2
+z_1\bar z_3}{1+z_3\bar z_3},\quad w_2=\frac{ z_2\bar z_3 -\bar
z_1}{1+z_3\bar z_3}. \label{wz}\eeq
 Choosing $z_3$ as a local
coordinate of  $S^2=\DC P^1$,
\beq u=z_3\;, \label{uz3}\eeq we
get the expressions
 \beq z_1=w_1u-\bar
w_2,\quad z_2=w_2u+\bar w_1,\quad z_3=u.
\label{bt}\eeq
In these coordinates the Fubini-Study metric
 on
$\DC P^3$ looks as follows:
\beq g_{a\bar b}dz_ad\bar z_b=\frac{dzd\bar z}{1+z\bar
z}- \frac{(\bar z dz)(zd\bar z)}{(1+z\bar z)^2}= \frac{dw_id\bar
w_i}{(1+w\bar w)^2}+ \frac{(du+{\cal A})(d\bar u+\bar{\cal
A})}{(1+u\bar u)^2},\label{map}\eeq
where \beq
{\cal A}=\frac{(\bar w_1+w_2u)(udw_1-d\bar w_2) +(\bar
w_2-w_1u)(udw_2+d\bar w_1)}{1+w\bar w}. \label{A}\eeq
Hence,
$w_1,w_2$ and $u$ are  the conformal-flat complex coordinates of
$S^4=\DH P^1$ and $S^2=\DC P^1$, while the connection ${\cal A}$
defines the $SU(2)$ gauge  field.\\

Now, let us
consider the Hamiltonian system describing the motion of a free particle on
$\DC P^3$
 \begin{equation}
{\cal H}_{\DC P^3}= g^{a \bar b}\pi_a{\bar \pi}_b \;,\quad
\{z_a,\pi_b\}=i\delta_{ab}\eeq
Let us extend   the coordinate transformation (\ref{bt}) to the $T^*\DC P^3$,
 by the
following transformation of momenta:
$$ \pi_1=\frac{\bar u
p_1-\bar p_2}{1+u\bar u},\quad \pi_2=\frac{\bar u p_2+
p_1}{1+u\bar u},$$
\beq
\pi_3= p_u+\frac{\bar  p_2 w_1-\bar p_1 w_2
- \bar u (w_1p_1+w_2p_2)}{1+u\bar u}. \label{pcan}\eeq
This extended transformation is a canonical
transformation,
\beq
 \{w_i,p_j\}=\delta_{ij},\quad \{u, p_u\}=1.
\label{htrans}\eeq
In the new coordinates
  the   Hamiltonian reads
\beq
 {\cal H}_{\DC P^3}=(1+w\bar w)^2P_i\bar P_i +(1+u\bar u)^2p_u\bar p_u\; .
 \eeq
Here we introduced
the covariant momenta
\beq P_1=p_1-i\frac{\bar w_1 }{1+w\bar
w}I_1-\frac{w_2}{1+w\bar w}I_+, \quad P_2=p_2-i\frac{\bar w_2
}{1+w\bar w}I_1+\frac{w_1}{1+w\bar w}I_+, \label{Pcov}\eeq
and
the $su(2)$ generators  $I_\pm, I_1$ defining the isometries of
$S^2$
\beq\begin{array}{c} I_1=-i(p_u u-\bar p_u \bar u), \quad
I_-=p_u+\bar u^2\bar p_{\bar u},
\quad  I_+=\bar p_{\bar u}+ u^2 p_u\\
\{I_\pm, I_1\}= \mp iI_\pm,\quad \{I_+, I_-\}= 2iI_1.
\end{array}
\label{Kiso}\eeq
 The nonvanishing Poisson brackets between $P_i$,
$w_i$ are given by the following relations (and their complex
conjugates):
 \beq \{w_i,P_j\}=\delta_{ij},\quad \{ P_1,
P_2\}=-\frac{2I_+}{(1+w\bar w)^2}, \quad\{ P_i,\bar
P_j\}=-i\frac{2I_1\delta_{ij}}{(1+w\bar w)^2}.
\label{instant}\eeq
The expressions in the r.h.s. define the strength of a homogeneous
$SU(2)$ instanton (the ``angular part" of the $SU(2)$ Yang monopole),
 written in terms of conformal-flat coordinates
of $S^4=\DH P^1$. Hence, the first part of the Hamiltonian, i.e.
${\cal D}_4=(1+w\bar w)^2P_i\bar P_i$, describes a
particle on the four-dimensional sphere in the field of a $SU(2)$ instanton.

The  Poisson brackets between $P_i$ and $u,\bar u, p_u,\bar p_u$
are  defined by the following nonzero relations and their complex
conjugates: \beq \{ P_i, p_u\}=-\frac{ \overline{{w}_{i}}+2
\epsilon_{ij}\, {w}_{j} \, u} {1+\overline{{w}}{w}}p_u,\;\; \{
P_i,\bar p_u\}=\frac{\overline{w}_i{\bar p_
u}}{1+\overline{w}{w}},  \;\; \nonumber\eeq \beq \left\{ P_{i},u\right\}
=\frac{ \left ( \overline{w_{i}}+\epsilon_{ij}w_{j} \, u\right )u
}{1+\overline{w} w}\;\;, \left\{ \overline{P_{i}},u\right\}
=\frac{ \epsilon_{ij}\overline{w_{j}}-w_{i} \,
u}{1+\overline{w}w}. \label{Ppbrac}\nonumber\eeq
The second part of the
Hamiltonian defines the motion of a free particle on the
two-sphere.
 It could be represented as a Casimir of $SU(2)$ \beq
{\cal D}_{S^2}=(1+u\bar u)^2p_u\bar p_u=I_+I_-+I_1^2\equiv I^2.
\label{D2}\eeq
It commutes with the Hamiltonian ${\cal D}_0$, as
well as with  $I_1, I_\pm$ and  $P_i, w_i$
\beq \{{\cal D}_{\DC
P^3}, I^2\}= \{P_i,{ I}^2\}_B= \{w_i,{ I}^2\}_B= \{{\cal I}_1,{
I}^2\}_B=\{{\cal I}_\pm ,{I}^2\}_B = 0.
\label{nen}\eeq
Hence, we
can perform a Hamiltonian reduction by the action of the generator
${\cal D}_2$, which reduces the initial twelve-dimensional phase
space $T_*\DC P^3=T^*(S^4\times S^2)$ to a ten-dimensional one.
 The relations
(\ref{nen}) allow us to  parameterize the reduced ten-dimensional
phase space in terms of the coordinates $P_i, w_i, I_\pm, I_1$,
where the latter obey the relation
\beq I_+I_- +I^2_1\equiv
I^2=const\;. \eeq
 Thus, the reduced phase space is nothing but  $T^*S^4 \times
S^2$, where $S^2$ is  the internal space of the instanton.

Let us collect the whole set of non-zero expressions  defining
the Poisson brackets on  $T_*S^4 \times S^2$
$$ \{w_i,
P_j\}=\delta_{ij},$$
$$\{ P_1, P_2\}=-\frac{2I_+}{(1+w\bar w)^2},$$
$$
 \{ P_i,\bar P_j\}=-i\frac{2I_1\delta_{ij}}{(1+w\bar w)^2},
$$
\beq\left\{ P_{i},I_{1}\right\}
=i\frac{\epsilon_{ij}w_{j} \, I_{+}}{1+ w\overline{w}} \label{Pb2}\eeq
$$
\left\{ P_{i},I_{+}\right \} =\frac{\overline{w_{i}} \, I_{+}}{1+
\overline{w} w}, $$
$$
 \left \{{P _{i}},I_{-}\right\} = -\frac{
\overline{w_{i}} \, I_{-}+ 2i\epsilon_{ij}{w_{j}}\, I_{1}
}{1+\overline{w} w} $$
$$ \{I_+,I_-\}=2iI_1,\quad
\{I_\pm, I_1\}=\mp iI_\pm \; .
$$
The reduced
Hamiltonian is $ {\cal H}_{\DC P^3}^{red}=(1+w\bar w)^2 P\bar P
+I^2.$
 So, the Hamiltonian of the  coloured particle
 on $S^4$ interacting with the $SU(2)$ instanton
is
connected with the Hamiltonian of a particle on $\DC P^3$ as
follows: \beq {\cal D}_{S^4}={\cal D}_{\DC P^3}^{\rm red}-I^2
\quad(>0). \eeq
This yields an intuitive explanation of the
degeneracy in the ground state in the corresponding quantum system
on
$S^4$. Indeed, since the l.h.s. is positive, the ground state of
the quantum system  on $S^4$ corresponds to the excited
state of a particle on $\DC P^3$, which is a degenerate one. On
the other hand, the ground state of a particle on $\DC P^3$ can be
reduced to the free particle on $S^4$, when $I=0$.
%

Now, let us consider a similar reduction for the particle on
$\DC P^3$, in the presence of
 constant magnetic field (\ref{ssB}).

Passing to the coordinates (\ref{bt}) and momenta (\ref{Pcov}) we
get the Poisson brackets defined by the nonzero relations given by
(\ref{Ppbrac}) and \beq
\{p_u, \overline{p}_u\}_B=\frac{iB}{(1+u\overline{u})^2},
\label{pupub}\eeq
\beq \{w_i, P_j\}_B=\delta_{ij},\quad \{ P_1,
P_2\}_B=-\frac{2{\cal I}_+}{(1+w\bar w)^2},
\eeq
\beq \{ P_i,\bar
P_j\}_B=-i\frac{2{\cal I}_1\delta_{ij}}{(1+w\bar w)^2}.
\label{instantB}\eeq where ${\cal I}_\pm, {\cal I}_1$ are defined
by the expressions \beq {\cal I}_1=I_1+\frac B2\frac{1-u\bar
u}{1+u\bar u}, \quad {\cal I}_-=I_- -B\frac{i\bar u}{1+u\bar u},
\quad  {\cal I}_+= I_+ +B\frac{i u}{1+u\bar u}\; \label{Kiso1}
\eeq
 Notice that the expressions (\ref{instantB}) are similar to
(\ref{instant}) and the generators (\ref{Kiso1})  form, with
respect to the new Poisson brackets, the $su(2)$ algebra \beq
\{{\cal I}_\pm, {\cal I}_1\}_B= \mp i{\cal I}_\pm,\quad \{{\cal
I}_+, {\cal I}_-\}= 2i{\cal I}_1. \eeq It is clear that these
generators define the isometries of the ``internal"
two-dimensional sphere  with a magnetic monopole located at the
center.

Once again, as in the absence of a magnetic field, we can reduce the
initial system by the Casimir of the $SU(2)$ group \beq {\cal
I}^2\equiv {\cal I}_1^2 + {\cal I}_+ {\cal I}_- ={\cal D}_{S^2}
+B^2/4,\quad \Rightarrow \quad {\cal I}\geq B/2\; .
\label{casB}\eeq In order to perform the Hamiltonian reduction, we
have to fix the value of ${\cal I}^2$, and then factorize by the
action of the vector field $\{{\cal I}^2,\;\}_B $.

The coordinates (\ref{wz}), (\ref{Pcov}) commute with the Casimir
(\ref{casB}), \beq \{P_i,{\cal I}^2\}_B= \{w_i,{\cal I}^2\}_B=
\{{\cal I}_1,{\cal I}^2\}_B=\{{\cal I}_\pm ,{\cal I}^2\}_B = 0.
\eeq Hence, as we did above, we can choose $P_i$, $w_i$, and
${\cal I}_\pm$ as the coordinates of the reduced, ten-dimensional
phase space.

The coordinates ${\cal I}_\pm$, ${\cal I}_1$ obey the condition
\beq {\cal I}^2_1+{\cal I}_-{\cal I}_+={\cal I}^2=const. \eeq The
resulting Poisson brackets are defined by the expressions (\ref{Pb2}),
with  $I_1, I_\pm$ replaced by ${\cal I}_\pm,{\cal I}_1$.

Hence, the particle on  $\DC P^3$ moving in the presence of a constant magnetic field
  reduces to a coloured particle on $S^4$ interacting with the instanton field.
  The Hamiltonians of these two systems are related as follows:
  \beq {\cal D}_{S^4}={\cal D}_{\DC P^3}^{red}-{\cal I}^2+B^2/4,\qquad
  {\cal I}\geq B/2
\eeq
Notice that, upon
quantization, we must replace ${\cal I}^2$ by ${\cal I}({\cal
I}+1)$ and require that both ${\cal I}$ and $B$  take
(half)integer values (since we assume unit radii for the spheres,
this means that the ``monopole number" obeys a Dirac quantization
rule).
The extension of this  reduction to quantum mechanics
relates the  theories of the quantum Hall effect on $S^4$ \cite{4h} and on $\DC P^3$
\cite{kn}.

Notice that the third Hopf map
could also be related with the generalized  quantum Hall effect theory \cite{Bernevig}.

\section{Generalized oscillators}
Among the integrable systems with hidden symmetries the oscillator is
the simplest one. In contrast to other systems with hidden symmetries
(e.g. Coulomb systems), its symmetries form a Lie algebra.
The $N$-dimensional oscillator on $T^*\DR^N$,
\beq
{\cal H}=\frac 12\left( p_ap_a+\alpha^2 q^aq^a\right),
\quad \omega_{can}=dp_a\wedge dq^a,
\qquad a=1,\ldots,N
\eeq
besides the rotational symmetry  $so(N)$, has also hidden ones,
so that the whole symmetry algebra is $su(N)$.
The symmetries of the oscillator are given by the generators
\beq
J_{ab}=p_aq^b-p_bq^a, \quad I_{ab}=p_ap_b+\alpha^2 q^a q^b.
\label{Josc}\eeq
The huge number  of hidden symmetries allows us to construct generalizations
of the oscillator on curved spaces, which inherit many properties
 of the initial system.

The  generalization of  the oscillator on the sphere was
suggested by Higgs \cite{higgs}.
It is given by the following Hamiltonian system:
\beq
{\cal H}=\frac 12 g^{ab}p_ap_a+\frac{\alpha^2}{2}q^aq^a\; ,
\quad \omega=dp_a\wedge dq^a,
\qquad  q^a=\frac{x_a}{x_0},
\label{Hhiggs}\eeq
where $x^a, x_0$ are the Euclidean coordinates of the ambient space $\DR^{N+1}$:
$ x^2_0+x^a x^a=1$, and $g_{ab}dq^adq^b$ is the metric on $S^N$.
This system inherits the rotational symmetries of the flat oscillator given by
(\ref{Josc}), and possesses the hidden symmetries given by the following
constants of motion (compare with (\ref{Josc})):
\beq
I_{ab}=J_aJ_b+\alpha^2 q^aq^b,
\label{Ihiggs}\eeq
where $J_a$ are the translation generators on $S^N$.

In contrast to the flat oscillator, whose symmetry algebra is $su(N)$,
the spherical (Higgs) oscillator has a nonlinear symmetry algebra.\\

This construction has been   extended to the complex projective
spaces in Ref. \cite{cpn}, where
 the oscillator  on $\DC P^N$ was defined by the Hamiltonian
\beq
{\cal H}={g^{\bar a b}\bar\pi_a\pi_b}+\alpha^2  z\bar z,
\label{Hoscpn}\eeq
with  $z^a=u^a/u^0$ denoting inhomogeneous coordinates of $\DC P^N$ and
$g_{a\bar b}dz^ad\bar  z^b$ being Fubini-Study metric (\ref{35}).

It is easy to see that this system has constants of motion
 given by the expressions
\beq
J_{a\bar b}={i}(z^b\pi_a-\bar\pi_b\bar z^a), \quad
I_{a\bar b}=
{J^+_a J^-_b} +\omega^2 {\bar z}^a z^b\; ,
\label{sym}\eeq
where $J^+_a=\pi_a+(\bar z\bar\pi)\bar z^a$, $J_a^-=\bar J^+_a$
are the translation generators on $\DC P^N$.
The generators $J_{a\bar b}$ define the kinematical symmetries of the system
and form a $su(N)$ algebra. When $N>1$, the generators $I_{a\bar b}$ are functionally
 independent of ${\cal H}$, $J_{a\bar b}$ and define hidden symmetries.
 As in the spherical case, their algebra is a nonlinear one
\beq
\begin{array}{c}
\{J_{{\bar a} b}, J_{\bar c d}\}=
i\delta_{\bar a d}J_{\bar b c}
-i\delta_{\bar c b}J_{\bar a d},\\
\{I_{a\bar b}, J_{c\bar d}\}=
i\delta_{c\bar b}I_{a\bar d}-i\delta_{a\bar d}I_{c\bar b} \\
 \{I_{a \bar b}, I_{c\bar d}\}=
i\alpha^2 \delta_{c\bar b} J_{a\bar d}- i\alpha^2
 \delta_{a\bar d}J_{c\bar b}+\\

+i I_{c\bar b}(J_{a\bar d}+J_0\delta_{a\bar d})-
i I_{a\bar d}(J_{c\bar b}+J_0\delta_{c\bar b}) \quad .
\end{array}
\label{cpnalg}\eeq
Hence, it is seen that for  $N=1$, i.e. in the case of the two-dimensional
sphere $S^2=\DC P^1$, the  suggested system has no hidden symmetries,
as  opposed to the Higgs oscillator on $S^2$.
Nevertheless, this model is exactly solvable both for $N=1$ and $N>1$ \cite{qcpn}.
Moreover, it remains exactly solvable, even after inclusion
of a constant magnetic field, for any $N$
(including $N=1$, when it has no hidden symmetries).
The magnetic field does not break the symmetry algebra of the system!
   As opposed to the described model,
    the constant magnetic field breaks
     the
     hidden symmetries, as well as the exact solvability,
      of the Higgs oscillator
       on $S^2=\DC P^1$.

{\bf Remark.}The Hamiltonian (\ref{Hoscpn}) could be represented
 as follows:
 \beq
{\cal H}=g^{a\bar b}(\pi_a{\bar\pi}_b+
\alpha^2\partial_a K {\bar\partial}_b K),
\label{Koscg}\eeq
where $K(z,\bar z)=\log (1+z\bar z)$
is the K\"ahler potential of the Fubini-Study metric.

Although this potential is not uniquely defined, it provides the system
with some properties, which
are general for the few oscillator models on K\"ahler spaces.
By this reason we postulate it as an oscillator potential on arbitrary
K\"ahler manifolds.\\

Now, let us compare these systems with the
 sequence which we like:
real, complex, quaternionic numbers (and zeroth, first, second Hopf map).
Let us observe, that the  $S^N$-oscillator potential is defined,
in terms of the ambient space   $\DR^{N+1}$,
in complete similarity to the $\DC P^N$-oscillator potential in terms of the
``ambient" space $\DC^{N+1}$.
The latter system  preserves its exact solvability
in the presence of a constant magnetic ($U(1)$ gauge) field.

Hence, continuing this sequence, one can define  on the quaternionic
projective
spaces $\DH P^N$ the oscillator-like system
given by the potential
\beq
V_{\DH P^N}=\alpha^2 w^a\bar w^a=
\alpha^2\frac{u^a_1\bar u^a_1 +u^a_2\bar u^a_2}{u^0_1 \bar u^0_1+ u^0_1 \bar u^0_1},
\eeq
where
$$
w^a=\frac{u^a_1+ ju^a_2}{u^0_1+ju^0_2},\quad
 u^a_1\bar u^a_1 +u^a_2\bar u^a_2+ u^0_1 \bar u^0_1+ u^0_2 \bar u^0_2=1\; .
$$
Here  $w^a$ are inhomogeneous (quaternionic)
coordinates of the quaternionic projective space $\DH P^N$,
and $u^a_0+ju^a_1, u^0_1+ju^0_2$ are  the Euclidean coordinates
of the ``ambient" quaternionic space $\DH^{N+1}=\DC^{2N+2}$.

One can expect that this system will be a superintegrable one
and will be exactly solvable also in the
presence of a $SU(2)$ instanton field.

In the simplest case of $\DH P^1= S^4$ we shall get the alternative (with respect to the Higgs)
model of the oscillator on the four-dimensional sphere.
In terms of the ambient space $\DR^5$, its potential will be given by the expression
\beq
V_{S^4}=\alpha^2\frac{1-x^0/x}{1+x^0/x}=\alpha^2\frac{1-\cos\theta}{1+\cos\theta}.
\eeq
Checking this system for this simplest case, we found,
  that it is indeed exactly solvable in the
  presence of the instanton  field \cite{mardoyan}.\\

 Let us mention that the Higgs (spherical) oscillator could be straightforwardly extended
 to (one- and two-sheet) hyperboloids,
  and the $\DC P^N$-oscillator - to the Lobachevsky spaces
  ${\cal L}_N=SU(N+1)/U(N)$. In both cases these systems  have hidden symmetries.

Notice also that, on the spheres $S^N$, there exists the
analog of the Coulomb system suggested by Schr\"odinger \cite{sphere}.
It is  given by the potential
\beq
V_{Coulomb}=-\frac{\gamma}{r_0}\frac{y_{N+1}}{|{\bf y}|},
 \qquad y^2_{N+1}+|{\bf y}|^2=r^2_0.
\eeq
This system  inherits the hidden symmetry of
the conventional Coulomb system on $\DR^N$.

Probably, as in the case of the oscillator,
one can define superintegrable analogs of the Coulomb system
on the complex projective spaces  $\DC P^N$ and on the quaternionic projective spaces
$\DH P^N$.
However, up to now, this question has not been analyzed.

\subsection*{Relation of the (pseudo)spherical oscillator and Coulomb systems}
The oscillator and Coulomb systems, being the best known among
the superintegrable mechanical systems,
 possess many similarities both at the classical
and quantum mechanical levels.
Writing down these systems in spherical coordinates,
one can  observe
that the radial Schr\"odinger equation of the $(p+1)$-dimensional Coulomb system could be transformed
in the Shr\"odinger equation of the $2p$-dimensional oscillator
 by the transformation (see, e.g. \cite{ta})
$$
r=R^2,
$$
where $r$ and $R$ are the radial coordinates
of the Coulomb and oscillator systems, respectively.

Due to the existence of the Hopf maps, in the cases of $p=1,2,4$ one can
establish a complete correspondence between these systems.
Indeed, their angular parts are, respectively,
$p$- and $(2p-1)$-dimensional spheres, while the above relation follows immediately
from (\ref{hm1}).
Considering
the Hamiltonian reductions related to the Hopf maps
(as it was done in the previous section),
 one can deduce,
that  the  $(p+1)$-dimensional
 Coulomb systems could be obtained from the $2p$- dimensional oscillator,
 by a reduction under the $G=S^{(p-1)}$ group. Moreover,
 for non-zero values of those generators we shall get
 generalizations of the Coulomb systems, specified by the presence of
 a magnetic flux ($p=1$), a Dirac monopole ($p=2$), a Yang monopole ($p=4$)
 \cite{ntt,ks,su2}.
However, this procedure assumes a change in the roles
of the coupling constants and the energy. To be more precise, these reductions
convert the energy surface  of the oscillator in the energy surface of the
Coulomb-like  system, while there is no one-to-one correspondence between
their Hamiltonians.

As we have seen above,
there exists well-defined  generalizations of
the oscillator systems on the spheres, hyperboloids, complex projective spaces
and Lobachevsky spaces.
The Coulomb system could also be generalized on the spheres and hyperboloids.
Hence, the following natural question arises.
Is it possible to relate the oscillator and Coulomb systems
on the spheres and hyperboloids, similarly to those in the flat cases?
The answer is positive, but it is rather strange.
The oscillators on the $2p$-dimensional
sphere and two-sheet hyperboloid (pseudosphere) result in the Coulomb-like systems
on the $(p+1)$-dimensional pseudosphere, for $p=1,2,4$ \cite{np}.

Below, following \cite{np}, we shall show how to relate the oscillator and Coulomb
systems on the spheres and two-sheet hyperboloids.
In the planar limit this relation results in the standard correspondence
between the conventional (flat) oscillator and the Coulomb-like system.
We shall discuss mainly the $p=1$ case,  since the treatment could be
 straightforwardly extended to the $p=2,4$ cases.

Let us introduce  the complex coordinate
$z$ parameterizing the sphere
by the complex projective plane $\DC P^1$ and
the two-sheeted hyperboloid by  the Poincar\'e
disk (Lobachevsky plane, pseudosphere) ${\cal L})$
\begin{equation}
{\bf x}\equiv x_1+ix_2=R_0\frac{2z}{1+\epsilon z\bar z},\quad
 x_3=R_0\frac{1-\epsilon {z\bar z}}{1+\epsilon {z\bar z}}.
\label{x}\end{equation}
In these coordinates   the metric becomes conformally-flat
\begin{equation}
 ds^2=R_0^2\frac{4dz d\bar z}{(1+\epsilon{z\bar z})^2}.
\label{met}\end{equation}
Here $\epsilon=1$ corresponds to the system on the sphere, and $\epsilon=-1$
to that on the pseudosphere.
The lower hemisphere and the lower sheet of the hyperboloid are
parameterized by the unit disk $|z|<1$, while the upper hemisphere
and the upper sheet of the hyperboloid are specified by
$|z|>1$, and transform
one into another  by the inversion $z\to 1/z$.
In the limit $R_0\to \infty$  the lower
 hemisphere (the lower sheet of the hyperboloid) turns into the
 whole two-dimensional plane.
 In these terms the oscillator and Coulomb potentials
read
\begin{equation}
V_{osc}=\frac{2\alpha^2 R_0^2z{\bar z}}{(1-\epsilon z{\bar z})^2},
\quad V_{C}=
-\frac{\gamma}{R_0}\frac{1-\epsilon z{\bar z}}{2|z|},
\label{v1}\end{equation}
Let us equip the  oscillator phase space
$T^*\DC P^1$ ($T^*{\cal L})$ with the symplectic structure
\begin{equation}
\omega=d\pi\wedge dz+ d{\bar\pi}\wedge d{\bar z}
\end{equation}
and  introduce the rotation generators
defining the $su(2)$ algebra for $\epsilon=1$
and the $su(1.1)$ algebra for $\epsilon=-1$
\begin{equation}
{\bf J}\equiv \frac{iJ_1 -J_2}{2}=\pi +\epsilon{\bar z}^2\bar\pi,
\;\; J\equiv \frac{\epsilon J_3}{2}=i(z\pi-{\bar z}{\bar\pi}).
\label{j}\end{equation}
These generators, together with ${\bf x}/R_0, x_3/R_0$,
define the algebra of motion of the (pseudo)sphere
 via  the following  non-vanishing Poisson brackets:
\begin{equation}
\begin{array}{c}
\{{\bf J}, {\bf x}\}=2x_3,\;\{{\bf J}, x_3\}=-\epsilon{\bf\bar x},\;
\{J, {\bf x}\}=i{\bf x}, \\
\{{\bf J}, {\bf\bar J}\}=-2i\epsilon J,\;
\{{\bf J}, J\}=i{\bf J}.
\end{array}
\label{alg}\end{equation}
In these terms,  the Hamiltonian of a free particle on the (pseudo)sphere
 reads
\begin{equation}
{ H}^{\epsilon}_0=
\frac{{\bf J}{\bf\bar J}+\epsilon J^2}{2R_0^2}
=\frac{(1+\epsilon z{\bar z})^2\pi{\bar\pi}}{2R_0^2},
\label{k}\end{equation}
whereas the oscillator Hamiltonian is given
 by the expression
\begin{equation}
{ H}^{\epsilon}_{osc}(\alpha,R_0|\pi, {\bar\pi}, z, {\bar z})=
\frac{(1+\epsilon z{\bar z})^2\pi{\bar\pi}}{2R_0^2}
+\frac{2\alpha^2R_0^2{z\bar z}}{(1-\epsilon{z\bar z})^2}.
\label{ho}\end{equation}
It can be easily verified that
the latter system  possesses the hidden symmetry given by the
complex (or vectorial) constant of motion \cite{higgs}
\begin{equation}
{\bf I}=I_1+iI_2=\frac{{\bf J}^2}{2R_0^2} +
\frac{\alpha^2R_0^2}{2}\frac{{\bf\bar  x}^2}{x^2_3},
\label{I}\end{equation}
which defines, together  with $J$ and $H_{osc}$ ,  the cubic algebra
\begin{equation}
 \{{\bf I}, J\}=2i{\bf I},\;\;\{{\bf\bar I},{\bf I}\}=4i
\left(\alpha^2 J +\frac{\epsilon JH_{osc}}{R_0^2}-\frac{J^3}{2R_0^4}\right).
\label{Ia}\end{equation}
The energy surface of the oscillator on the (pseudo)sphere
$H^\epsilon_{osc}=E$  reads
\begin{equation}
\frac{\left(1-(z\bar z)^2\right)^2{\pi}{\bar\pi}}{2R_0^4}+
2\left(\alpha^2+\epsilon \frac{E}{R^2_0}\right) z{\bar z}
=\frac{E}{R^2_0}\left(1+(z\bar z)^2\right).
\label{os}\end{equation}
Now, performing the canonical Bohlin transformation (\ref{b})
one can rewrite the expression (\ref{os})
as follows:
\begin{equation}
\frac{(1-w\bar w)^2{p}{\bar p}}{2r_0^2}-
\frac{\gamma}{r_0}\frac{1+w{\bar w}}{2|w|}={\cal E}_{C},
\label{C}\end{equation}
where we introduced the notation
\begin{equation}
r_0=R_0^2,\quad  \gamma=\frac{E}{2},
\quad -2{\cal E}_{C}=\alpha^2+\epsilon\frac{E}{r_0}.
\label{gam}\end{equation}
Comparing the l.h.s. of (\ref{C}) with the expressions (\ref{v1}), (\ref{k})
we  conclude that (\ref{C}) defines the energy surface
of the Coulomb system
 on the pseudosphere with ``radius'' $r_0$,
 where  $w, p$ denote the complex
 stereographic coordinate and its conjugated momentum, respectively.
In the above, $r_0$ is the ``radius'' of the pseudosphere, while ${\cal E}_C$ is the
 energy of the system.
{\it Hence, we related
classical isotropic oscillators on the sphere and pseudosphere
with the  classical Coulomb problem on the pseudosphere.}

The constants of motion of the oscillators, $J$ and ${\bf I}$
(which   coincide
 on the energy surfaces (\ref{os})) are  converted, respectively,
 into the doubled
angular momentum and the doubled Runge-Lenz vector
 of the Coulomb system
\begin{equation}
J\to 2J_{C},\quad
 {\bf I}\to 2{\bf A},\quad {\bf A}=
-\frac{iJ_{C}{\bf J}_{C}}{r_0}+
{\gamma}\frac{{\bf\bar x}_{C}}{|{\bf x}_{C}|},
\end{equation}
where ${\bf J}_{C}$, $J_{C}$, ${\bf x}_{C}$ denote the rotation generators
and  the pseudo-Euclidean coordinates  of the Coulomb system.

 We have shown above that, for establishing the quantum-mechanical
 correspondence, we have to supplement the quantum-mechanical
 Bohlin  transformation
with the reduction by the $Z_2$ group action,
 choosing either
 even  ($\sigma=0$) or odd ($\sigma=1/2$) wave functions (\ref{3b}).
 The resulting Coulomb system
 is  spinless for $\sigma=0$, and it possesses  spin $1/2$ for $\sigma=1/2$.

The presented construction could be straightforwardly extended to higher dimensions,
concerning the $2p-$dimensional oscillator on the (pseudo)sphere
and  the $(p+1)-$dimensional
Coulomb-like systems,  $p=2,4$.
It is clear, that the $p=2$ case corresponds
to the Hamiltonian reduction, associated with the first Hopf map, and the $p=4$
case is related to the second Hopf map.
Indeed,   the
oscillator on the $2p$-dimensional
(pseudo)sphere  is also described  by the
Hamiltonian (\ref{ho}),
 where   the following replacement is performed:
$(z,\;\pi)\to (z^a,\;\pi_a)$, $a=1,\ldots, p$,
with the summation over these indices understood.
Consequently,  the oscillator energy surfaces  are again given by Eq. (\ref{os}).
Then, performing the Hamiltonian reduction, associated with the $p$-th Hopf maps
(see the previous Section) we shall get the  Coulomb-like
system on the $(p+1)$-dimensional pseudosphere.\\

For example, if $p=2$, we reduce  the system under consideration
by the Hamiltonian action of the $U(1)$ group given by the generator
$J=i(z\pi-{\bar z}{\bar\pi})$. This reduction was described in detail in
 {\sl Section 2}.
For this purpose, we have to  fix the level surface $J=2s$
and choose the $U(1)$-invariant  stereographic coordinates
in the form of the  conventional
Kustaanheimo-Stiefel transformation (\ref{ksc}).
The resulting symplectic structure takes the form (\ref{18}).
The oscillator energy surface
reads
\begin{equation}
\frac{(1- {\bf q}^2)^2}{8r_0^2}({\bf p}^2 +\frac{s^2}{{\bf q}^2})-
\frac{\gamma}{r_0}\frac{1+{\bf q}^2}{2|{\bf q}|}={\cal E}_C,
\label{C3}\end{equation}
where  $r_0$, $\gamma$, ${\cal E}_C$
are defined by the expressions (\ref{gam}).

Interpreting ${\bf q}$  as the
 (real) stereographic coordinates of the three-dimensional
pseudosphere
\begin{equation}
{\bf x}=r_0\frac{2{\bf q}}{1- {\bf q}^2},\quad
 x_4=r_0\frac{1+ {\bf q}^2}{1- {\bf q}^2},
\label{hyper3}
\end{equation}
  we conclude that (\ref{C3}) defines  the energy surface of
the  pseudospherical analog of a Coulomb-like system
proposed in Ref. \cite{Z}, which is also known
under the name of ``MIC-Kepler" system.\\

In the $p=4$ case, we have to reduce the system by  the  action
of the $SU(2)$ group
and choose the $SU(2)$-invariant stereographic coordinates and momenta
in the form  corresponding to the standard Hurwitz transformation,
which yields a pseudospherical analog of the so-called $SU(2)$-Kepler
(or Yang-Coulomb) system  \cite{su2}.
The potential term of the resulting system will be given by the expression
\begin{equation}
V_{SU(2)-Kepler}=\frac{I^2}{r_0^2}\left(\frac{x^2_5}{2{\bf x}^2}-2\right)-
\frac{\gamma}{2r_0}\frac{x_5}{|{\bf x}|}
\label{C31}\end{equation}
where   $({\bf x}, x_5)$ are the (pseudo)Euclidean  coordinates of
the ambient space $\DR^{1.5}$ of the five-dimensional hyperboloid, $|{\bf x}|^2-x^2_5$;
 $I^2$ is  the value of the generator
 ${\cal J}_i^2$, under which
 the $SU(2)$ reduction has been performed.
 The constants   $r_0$, $\gamma$
 are defined by the expressions  (\ref{gam}).\\

 It is interesting to clarify, which systems will the $\DC P^N$-oscillators,
 after similar reductions, result in.
 We have checked it only for the first Hopf map, corresponding to the case
 $p=2$ \cite{bny,cpn}.To our surprise, we found that
 the oscillators on $\DC P^2$
 and ${\cal L}_2$ also resulted, after reduction, in the pseudospherical
 MIC-Kepler system!

\section{Supersymplectic structures}
In the previous Sections we presented some elements of Hamiltonian
formalism which, in our belief, could be useful in the study of
supersymmetric mechanics.

In the present Section we shall briefly discuss the Hamiltonian
formalism on superspaces (super-Hamiltonian formalism).
The super-Hamiltonian formalism, in its main lines, is a straightforward
extension of the ordinary Hamiltonian formalism to  superspace,
with a more or less obvious placement of sign factors. Probably, from the supergeometrical
viewpoint,  the only qualitative difference  appears in the existence of the odd Poisson brackets
(antibrackets), which have no analogs in ordinary spaces,
 and in the respect  of the differential forms to integration.
 Fortunately, these aspects
 are inessential for our purposes.\\

The Poisson brackets of the functions
 $f(x)$ and $g(x)$ on superspaces
 are defined by the expression
 \begin{equation}
\label{eq:bloc}\{f,g\}_\kappa=\frac{\partial _rf}{\partial x^A}\Omega _{\kappa}^{AB}(x)
\frac{\partial _lg}{\partial x^B},\qquad \kappa=0,1.
\end{equation}
They  obey the conditions
\begin{eqnarray}
& &p(\{ f, g \}_{\kappa} )= p(f)+ p(g) + \kappa
  \quad {\rm (grading )} ,
\nonumber \\
& &\{ f, g \}_\kappa = -(-1)^{(p(f)+\kappa )(p(g)+\kappa)}\{ g, f \}_\kappa
\quad {\rm ( "antisymmetricity")} ,\label{eq:anti} \\
& & ( -1)^{(p(f)+1)(p(h)+\kappa)}\{ f,\{ g, h \}_{\kappa}\}_\kappa +
{\rm {cycl. perm. (f, g, h)}} = 0
  \quad{\rm {(Jacobi\quad id.)}} .
\label{eq:bjac} \end{eqnarray}
Here $x^A$ are  local coordinates of superspace, while
$\frac{\partial _r}{\partial x^A} $ and $\frac{\partial _l}{\partial x^A}$
denote right and left derivatives, respectively.

It is seen that the nondegenerate
odd Poisson brackets can be defined on the $(N.N)$-dimensional
superspaces, and the nondegenerate even Poisson brackets could be defined
on the $(2N.M)$-dimensional ones.
In this  case  the Poisson brackets are associated with the
supersymplectic structure
\begin{equation}
\label{eq:symp}\Omega _\kappa=dz^A\Omega _{(\kappa )AB}dz^B,\quad d\Omega_\kappa=0
\end{equation}
where $\Omega _{(\kappa )AB}\Omega _{\kappa}^{BC}=\delta _A^C$.\\

 The generalization of the Darboux theorem states that locally, the
nondegenerate  Poisson brackets could be  transformed to the canonical form.
The canonical odd Poisson brackets look as follows:
\begin{equation}
\label{eq:bcan}\{f,g\}_1^{{\rm can}}=\sum_{i=1}^N\left( \frac{\partial _rf}{%
\partial x^i}\frac{\partial _lg}{\partial \theta _i}-\frac{\partial _rf}{%
\partial \theta _i}\frac{\partial _lg}{\partial x^i}\right) ,
\end{equation}
where $p(\theta_i)=p(x^i)+1=1$.
The canonical even Poisson brackets read
   \begin{equation}
          \{f,g\}_0
             =
            \sum_{i=1}^N \left(
           \frac{\partial f}{\partial  x^i}
           \frac{\partial g}{\partial x^{i+N} }
               -
        \frac{\partial f}{\partial  x^{i+N}}
        \frac{\partial g}{\partial  x^i}
               \right)
               +
              \sum_{\alpha =1}^{M}
        \epsilon _\alpha \frac{\partial_{r} f}{\partial  \theta^ \alpha}
        \frac{\partial_l{L} g }{\partial  \theta^{\alpha}} ,
        \;\;\epsilon_{\alpha} =\pm 1\; .
\end{equation}
Here  $x^i, x^{i+N}$ denote even coordinates, $p(x)=0$, and $\theta^\alpha$
are the odd ones $p(\theta)=1$.

In a completely similar way to the ordinary (non-``super") space, one can show
that the vector field preserving the supersymplectic structure
is a locally Hamiltonian one.
Hence,  both types of supersymplectic structures  can be related with the Hamiltonian
systems, which have the following equations of motion:
\beq
\frac{dx^A}{dt}=\{{\cal H}_\kappa ,x^A\}_\kappa ,\qquad p({\cal H}_\kappa )=\kappa.
\eeq
Any supermanifold ${\cal M}$
underlied by the bosonic manifold $M_0$ can be associated with
some vector bundle $VM_0$ of $M_0$ \cite{berezin}, in the following sense.
One can  choose on ${\cal M}$ local coordinates $(x^i,\theta^\mu)$, such
that the transition functions from one chart (parameterized by $(x^i,\theta^\mu)$)
to the other chart (parameterized by $({\tilde x}^i,{\tilde\theta}^\mu)$)
look as follows:
\beq
{\tilde x}^i={\tilde x}^i(x),\qquad {\tilde\theta^\mu}=A^{\mu}_{\nu}(x)\theta^{\nu}.
\label{LT}\eeq
Changing the parity of $\theta$: $p(\theta^\mu)=1 \to p(\theta^\mu )=0$,
we shall get the vector bundle $VM_0$ of  $M_0$.\\

Any supermanifold equipped with the odd symplectic structure,
is associated with the cotangent bundle of $M_0$ \cite{bschwarz},
so that the odd symplectic structure could be globally transformed
to the canonical form, with the odd Poisson bracket given by the expression
 (\ref{eq:bcan}). Hence, the functions on the odd symplectic manifold
 could be interpreted as  contravariant antisymmetric tensors on $M_0$.

 The structure of the even symplectic manifold is not so rigid: there is a
  variety of ways to extend the given  symplectic manifold $( {M}_0, \omega)$
  to the supersymplectic ones, associated with the vector bundle $VM_0$.
On these supermanifolds one can (globally) define the even symplectic structure
\begin{eqnarray}
 \label{tO}
 \Omega
 &=&\omega+
 d\left(\theta^{\mu}g_{\mu\nu}(x){\cal D}\theta^{\nu}\right)
 \nonumber\\
 &=&
 \omega + \frac{1}{2}
 R_{\nu \mu ki}\theta^{\nu}\theta^{\mu}dx^i\wedge dx^k
 +g_{\mu\nu}{\cal D}
 \theta^{\nu} \wedge {\cal D}\theta^{\mu}\,,
\end{eqnarray}
Here $x^i$ are local coordinates of ${\cal M }_0$ and
$\theta^\mu$ are the (odd) coordinates in the bundle;
   $g_{\mu\nu}=g_{\mu\nu}(x)$ are the components of the metrics in the bundle,
while ${\cal D}\theta^\mu=d\theta^\mu+\Gamma^\mu_{\;\nu
i}\,\theta^\nu dx^i$, where
  $\Gamma^\mu_{\;i\nu}$ are the connection components respecting the metric in the bundle
\beq
 g_{\mu\nu;k} = g_{\mu\nu,k} - g_{\mu\alpha}\Gamma^{\alpha}_{k\nu} -
 g_{\alpha\nu}\Gamma^{\alpha}_{k\mu} = 0\,.
 \label{gG}
\eeq
We used the following notation as well:
 $R_{\mu \nu
ki}=g_{\mu\alpha}R^{\alpha}_{\;\nu ki}$, where  $R^{\mu}_{\;\nu
ki}$ are  the components of connection's curvature
$$
 R^{\nu}_{~\alpha ki}
 =-\Gamma^{\nu}_{~k\alpha ,i}+
 \Gamma^{\nu}_{~i\alpha ,k} +\Gamma^{\nu}_{~k\beta}\Gamma^{\beta}_{~i\alpha}
 -
 \Gamma^{\nu}_{~i\beta}\Gamma^{\beta}_{~k\alpha} \,;\quad
 R^{\nu}_{~\alpha i k}
 =-R^{\nu}_{~\alpha k i}\;.
$$
Let us consider the coordinate transformation (\ref{LT}).
With respect to this transformation, the connection components transform as follows:
\begin{eqnarray}
\label{Gtrn}
 {\bar \Gamma}^{\mu}_{~i\nu}
 =A^{\mu}_{\;\;\lambda}\Gamma^{\lambda}_{~k\alpha}
 \frac{\partial_r x^k}{\partial {\bar x}^i}B^{\alpha}_{\;\;\nu}-
 A^{\mu}_{\;\;\alpha ,k}B^{\alpha}_{\;\;\nu}
 \frac{\partial_r x^k}{\partial {\bar x}^i}\, ,
 \qquad  A_{\mu}^{\;\;\nu}B_{\nu}^{\;\;\lambda}=
 \delta_{\mu}^{\lambda}.
\end{eqnarray}
Since  ${\cal D}\theta^{\nu}$ transforms homogeneously
under (\ref{LT}),
${\cal D}{\bar \theta}^{\nu}={\cal D}\theta^{\mu}A_{\mu}^{\;\;\nu}(x)$,
we conclude that the supersymplectic structure
(\ref{tO})
is covariant under (\ref{LT}) as well.

The corresponding Poisson brackets look as follows:
\beq
 \{f, g\}
 =(\nabla_i f)\,{\widetilde\omega}^{ij}( \nabla_j g)
 + \alpha\,\frac{\partial_r f}{\partial \theta^\mu}\,
 g^{\mu\nu}\frac{\partial_l g}{\partial \theta^\nu}\,;
\label{sssuperman} \eeq
where
$${\widetilde\omega}^{im}
 ({\omega}_{mj}+\frac{1}{2}
 R_{\nu \mu mj}\theta^{\nu}\theta^{\mu})=\delta^i_j,\qquad\nabla_i=
 \frac{\partial}{\partial x^i}
 - \Gamma^k_{\,ij}(x)\,\theta^{ja}\frac{\partial}{\partial\theta^{ka}}.$$

On the supermanifolds one can  define also  the analog of the
K\"ahler structures.
We shall call the complex symplectic supermanifold an
even (odd)    K\"ahler one,
when the even (odd) symplectic
    structure is defined by
the  expression
 \begin{equation}
     \Omega_{\kappa}=i(-1)^{p_A(p_B+\kappa+1)}g_{(\kappa )A {\bar B}}
           dz^A \wedge d{\bar z}^B,
\end{equation}
      where
$$  g_{(\kappa )A {\bar B}}  =
     (-1)^{(p_A+\kappa+1)(p_B+\kappa+1)+\kappa +1}
         \overline {g_{(\kappa )B {\bar A}}},\quad p(g_{(\kappa )A\bar B})=p_A
+p_B+\kappa .$$
Here and in the following, the index $\kappa =0(1)$ denotes the even(odd) case.

The K\"ahler potential on the supermanifold is a  local real
even (odd) function $K_\kappa (z,{\bar z})$ defining the K\"ahler structure
\begin{equation}
            g_{(\kappa )A {\bar B}}  =
              \frac{\partial_l}{\partial z^A}
        \frac{\partial_r}{\partial {\bar z}^B}
                   K_\kappa  (z,{\bar z}).
\end{equation}
As in the usual case, $K_\kappa$ is
   defined  up to arbitrary holomorphic and antiholomorphic  functions.

    With the even (odd) form $\Omega_{\kappa}$ one can associate
        the even (odd) Poisson bracket
   \begin{equation}
             \{ f,g\}_\kappa
                        =
             i\left(
         \frac{\partial_r f}{\partial \bar z^A}
               g^{(\kappa){\bar A}B}
            \frac{\partial_l g}{\partial z^B}
                   -
            (-1)^{(p_A+\kappa)(p_B+\kappa)}
             \frac{\partial_r f}{\partial z^A}
             g^{(\kappa ){\bar A}B}
         \frac{\partial_l g }{\partial \bar z^B}
               \right),
\end{equation}
 where
$$g^{(\kappa){\bar A}B} g_{(\kappa)B{\bar C}}=\delta^{\bar A}_{\bar C}
\;\;,\;\;\;\; \overline{g^{(\kappa){\bar A}B}}
 = (-1)^{(p_A+\kappa)(p_B+\kappa)}g^{(\kappa ){\bar B}A} .$$
{\bf Example.} Let us consider the supermanifold $\Lambda M$
associated with the tangent bundle
of the K\"ahler manifold $M_0$. On this supermanifold one can define the
even and odd K\"ahler potentials\cite{tmp}
\beq
K_0=K(z,\bar z)+F(ig_{a\bar b}\sigma^a\bar\sigma^b),\quad
K_1=\frac{\partial K(z,\bar z)}{\partial z^a}\sigma^a +
\frac{\partial K(z,\bar z)}{\partial{\bar z}^a}\bar\sigma^a
,
\eeq
where $K(z,\bar z)$ is a K\"ahler potential on $M_0$,
$g_{a\bar b}=\partial^2 K/\partial z^a\partial \bar z^b$, and $F(x)$ is
a real  function which obeys the condition $F'(0)\neq 0$.
It is clear that these functions define even and odd
K\"ahler structures on $\Lambda M_0$, respectively.\\

Finally, let us notice that
the analog of the Liouville measure for the even supersymplectic
symplectic
structure $\Omega_0$
reads
\begin{equation}
       \rho=\sqrt{{\rm Ber} \Omega_{(0) AB}},
\end{equation}
while the odd symplectic structure has no similar invariant \cite{kts}.
Indeed, one can verify that the even super-Hamiltonian vector field
is always divergenceless, ${\rm str} \{H,\;\}_0=0$
(similarly to the non-superHamiltonian vector field),
while in the case of the odd super-Hamiltonian vector field
this property of the Hamiltonian vector field fails.
As a consequence, in the latter case the so-called $\Delta$-operator can be defined
\cite{K}, which plays a crucial role in the
Batalin-Vilkovisky formalism (Lagrangian BRST quantization formalism) \cite{bv}.

\subsection*{Odd super-Hamiltonian mechanics}
Let us consider the supermanifold $\Lambda M$, associated with the tangent bundle of the
symplectic manifold $(M,\omega )$,  i.e.  the external algebra of
$(M,\omega )$.
In other words, the odd coordinates   $\theta ^i$ transform from one
chart to another
 like $dx^i$,  and they can be interpreted as the basis of the 1-forms on $M$.
 By the use of the $\omega$ we can equip $\Lambda M$
with the odd symplectic structure
\begin{equation}
\label{eq:osym}\Omega_{1} =d\left(\omega_{ij}\theta^j dx^i\right)=
\omega_{ij}dx^i\wedge d\theta^j
+\frac 12 \omega_{ki,j}\theta^j dx^k\wedge dx^i.
\end{equation}
The corresponding odd Poisson brackets are defined by the following
relations:
\begin{equation}
\label{eq:bxt}\{x^i,x^j\}_1=0,\quad \{x^i,\theta ^j\}_1=\omega ^{ij},\quad
\{\theta ^i,\theta ^j\}_1=\frac{\partial \omega ^{ij}}{\partial x^k}\theta
^k,
\end{equation}
where $\omega ^{ij}\omega_{jk}=\delta^i_k$.

Let us define, on ${\Lambda M}$,  the even function
\begin{equation}
\label{eq:F0}F=-\frac 12\theta^i\omega _{ij}\theta ^j,\; : \quad \{F,F\}_1=0,
\end{equation}
where the latter equation holds  due to the closeness  of $\omega$.
By making use  of this function, one can define the  map of any
function on   $M$ in the odd function on ${\Lambda M}$
\begin{equation}
\label{eq:map}f(x)\to Q_f(x,\theta )=\{f(x),F(x,\theta )\}_1\; ,
\end{equation}
which  possesses  the following important property:
\begin{equation}
\label{eq:cons}\{f(x),g(x)\}=\{f(x),Q_g(x,\theta )\}_1\quad {\rm for}\quad
{\rm any}\quad f(x),g(x) .
\end{equation}
In particular,  (\ref{eq:map})  maps the Hamiltonian mechanics
$\left( M,\omega, H(x)\right)$ in the following
super-Hamiltonian one: $\left( \Lambda M,\Omega _1, Q_H=\{H,F\}_1\right)$,
 where $Q_H$ plays the role of the odd Hamiltonian on ${\Lambda M}$.

The functions
 $H,F,Q_H$
 form the
superalgebra
\begin{equation}
\begin{array}{c}
  \{H\pm F, H \pm F \}_{1} =\pm 2Q_{H} ,\\
  \{H+ F, H -  F\}_{1} = \{H\pm F, Q_{H} \}_{1}
= \{Q_{H}, Q_{H} \}_{1} = 0,
\end{array}
\label{eq:sualg2}
 \end{equation}
i.e. the resulting mechanics possesses the supersymmetry transformation
defined by the ``supercharge"  $H+F$.
This  superalgebra  has a transparent interpretation in terms
of base  manifold $(M,\omega)$
$$
  \{H,\quad\}_{1}=\xi_{H}^i \frac{\partial}{\partial \theta^i}
\rightarrow \;\hat\imath _{H} -{\rm contraction} \;{\rm with}\; \xi_{H} ,
$$
$$
\{F,\quad\}_{1}= \theta^i
 \frac{\partial}{\partial x^i}  \; \rightarrow \; \hat d -
{\rm exterior\quad differential},
$$
$$
\{ Q,\quad \}_{1}=\xi_{H}^i
 \frac{\partial}{\partial x^i} + \xi_{H,k}^{i}\theta^k
\frac{\partial}{\partial \theta^i} \rightarrow  \hat{\cal L}_{H} -{\rm
Lie\quad derivative\quad along}\quad \xi_{H} ,
$$
 while, using the Jacobi identity (\ref{eq:bjac}), we
get
\begin{equation} \{H,F\}_1=Q_H\rightarrow \hat d\hat \imath _H+\hat
\imath _H\hat d=\hat{\cal L}_H-{\rm homotopy\quad formula}.\nonumber
\end{equation}
Hence, the above dynamics could be useful for the description of the
differential calculus on the symplectic (and Poisson) manifolds.
Particulary, it has a nice application in equivariant cohomology and
related localization formulae (see \cite{lnp} and refs therein).

However, the presented supersymmetric model has no  deep dynamical meaning,
since the odd Poisson brackets do not admit any consistent quantization scheme.
Naively, this is reflected in the fact that conjugated operators should
have opposite Grassmann grading, so that the Planck constant
must be a Grassmann-odd number.

Moreover, the presented supersymmetric mechanics
is not interesting even from the classical viewpoint.
Its equations of motion
read
$$\frac{dx^i}{dt}=\{x^i,Q_H\}_1=\xi _H^i,\quad\quad\quad
\\\frac{%
d\theta ^i}{dt}=\{\theta ^i,Q_H\}_1=\frac{\partial \xi _H^i}{\partial x^j}%
\theta ^j,
$$
i.e. the ``fermionic" degrees of freedom have no impact in the
dynamics of the ``bosonic" degrees of freedom.

Nevertheless, the odd Poisson brackets  are widely known, since 1981,
 in the theoretical physics community  under the name of ``antibrackets".
 That was the year, when Batalin and Vilkovisky suggested their {\sl Covariant
 Lagrangian BRST quantization formalism} (which is known presently as
 the Batalin-Vilkovisky  formalism) \cite{bv}, where the
  antibrackets (odd Poisson brackets) play the  key role.
  However, only decades after, this
  elegant formalism was understood in terms of conventional
  supergeometrical constructions \cite{K,bschwarz}.
It seems that the Batalin-Vilkovisky formalism
could also be useful for the geometrical (covariant)
formulation of the superfield approach to the construction of supersymmetric
Lagrangian field-theoretical and mechanical models \cite{aksz}.

We shall not touch upon these aspects of super-Hamiltonian systems,
and  will restrict ourselves to the consideration of supersymmetric
Hamiltonian systems with even symplectic structure.

\subsection*{Hamiltonian reduction: $\DC^{N+1.M}\to\DC P^{N.M}$,
 $\Lambda\DC^{N+1}\to \Lambda\DC P^N$}
The procedure of super-Hamiltonian reduction is very similar
to the Hamiltonian one.
The main difference is in the counting of the dimensionality of the phase superspace.
Namely, we should separately count the number of
``fermionic" and ``bosonic" degrees of freedom,
which were eliminated during the reduction.

Instead of describing the extension of the Hamiltonian reduction  to the supercase,
we shall  illustrate it by considering superextensions of the reduction
$\DC^{N+1}\to\DC P^N$  presented in Third Section. These examples were
considered in details in Ref. \cite{jmp2}.\\

Let us consider the complex superspace $\DC ^{N+1,M}$ parameterized by the complex
  coordinates $(u^{\tilde a}, \eta^n )$, ${\tilde a}=0,1,...,N$, $n=1,\ldots, M$.
Let us equip it with the canonical symplectic structure
       $$
      \Omega^0 = i(du^{\tilde a} \wedge {\bar du}^{\tilde a}
      -i d\eta^n \wedge d{\bar \eta}^ n)
$$
 and with the  corresponding even Poisson bracket
\begin{equation}
   \{ f,g\}_0 = i\left(\frac{\partial f}{\partial u^{\tilde a}}
                         \frac{\partial g}{\partial {\bar u}^{\tilde a}}    -
                        \frac{\partial f}{\partial {\bar u}^{\tilde a}}
                        \frac{\partial g}{\partial u^{\tilde a}}\right)   +
          \frac{\partial_r f}{\partial \eta^n}
          \frac{\partial_l g}{\partial {\bar \eta}^n} +
         \frac{\partial_r f}{\partial {\bar \eta}^n}
          \frac{\partial_l g }{\partial  \eta^n}.
\end{equation}
The (super-)Hamiltonian action of the $U(1)$ group is given, on this space, by the
generator
 \begin{equation}
        {\cal J}_0= u^{\tilde a} {\bar u}^{\tilde a} -i \eta^n {\bar \eta}^n.
\label{sJ}\end{equation}
For the reduction of $\DC^{N+1.M}$ by this generator,
we have to factorize the
    $(2N+1.2M)_{\DR}$-dimensional level supersurface
\begin{equation}
                         {\cal J }_0={ r}^2_0
\end{equation}
    by the even super-Hamiltonian vector field $\{{\cal J}_0,\;\}$
    (which is tangent to that surface). Hence, the resulting phase superspace is
    a $(2N.2M)_{\DR}$-dimensional one.

Hence, for the role of local coordinates of the reduced phase
space, we have to choose the $N$ even and $M$ odd complex  functions
 commuting with ${\cal J}_0$.
On the chart $u^{\tilde a} \neq 0$,  appropriate functions are the
following ones:
\begin{equation}
  z^A_{({\tilde a})}=\left( z^a_{({\tilde a})}=\frac{u^a}{u^{\tilde a}}\;,
  \;\theta^k_{({\tilde a})}=\frac{\eta^k}{u^{\tilde a}},
  \quad a \neq {\tilde a}\;\right) \;:
\{ z^A_{ ({\tilde a}) } , {\cal J }_0 \}_0=0\;.
\label{szt}\end{equation}
 The reduced Poisson brackets could be
  defined  by the expression $\{f,g\}_0^{\rm red}=
               \{f,g\}_0\mid_{{\cal J}_0={r}^2_0}$,
    where $f,g$ are functions depending on the coordinates
    $z^A_{({\tilde a})}, {\bar z^A_{({\tilde b})}}$.
    Straightforward calculations yield the result
             $$
\{ z^A,z^{ B}\}_0^{\rm red} =\{{\bar z}^{ A},{\bar w}^{ B}\}_0^{\rm red} = 0,
$$
$$
\{z^A,{\bar z}^{ B}\}_0^{\rm red}=
 (i)^{p_A p_B +1} \frac{1+(-i)^{p_C} z^C {\bar z}^{ C}}{{r}^2_0}
         \left(\delta^{AB}+(-i)^{p_A p_B} z^A {\bar w}^{B}\right).
         $$
It is seen that these Poisson brackets are associated with a K\"ahler structure.
This K\"ahler structure  is defined by the potential
              \beq
      K= { r}^2_0\log (1+(-i)^{p_C} z^C{\bar z}^{\bar  C}).
              \eeq
The transition functions from
the ${\tilde a}$-th chart to the ${\tilde b}$-th one look as follows:
\begin{equation}
        z^{\tilde c}_{({\tilde a})} =
        \frac{z^{\tilde c}_{({\tilde b})}}{z^{\tilde a}_{({\tilde b})}},  \;\;
       \quad \theta^k_{({\tilde a})} =
          \frac{\theta^k _{({\tilde b})}}{z^{\tilde a}_{({\tilde b})}},
           \quad {\rm where}\quad z^{\tilde a}_{({\tilde b})}=
     \left( w^{a}_{({\tilde b})},\; w^{\tilde a}_{({\tilde a})}=1 \right).
\end{equation}
Upon these transformations the K\"ahler potential changes on
the holomorphic and anti-holomorphic
functions, i.e. the reduced  phase space is indeed a K\"ahler supermanifold.
We shall refer to it as $\DC P^{N.M}$.
The quantization of this supermanifold is considered in
\cite{Ivanov}.\\

Now, let us consider the Hamiltonian reduction of the superspace
 $\DC^{N+1,N+1}$ by the action of the ${\cal N}=2$ superalgebra, given by the generators
\beq\begin{array}{c}
{\cal J}_0= u^{\tilde a} {\bar u}^{\tilde a} -i \eta^{\tilde a} {\bar \eta}^{\tilde a},
\quad
 \Theta^+=u^{\tilde a}{\bar\eta}^{\tilde a},
 \quad \Theta^-={\bar u}^{\tilde a}\eta^{\tilde a}\;:\\
 \{\Theta^+,\Theta^-\}={\cal J}_0,\;
 \{\Theta^{\pm},\Theta^{\pm}\}=\{\Theta^\pm ,{\cal J}_0\}=0.
\end{array}
\eeq
    The equations
  \begin{equation}
     {J}_0=r^2_0, \;\;\;\;   \Theta^\pm =0
\end{equation}
   define the $(2N+1.2N)$-dimensional level surface
   $M_{r^2_0,0,0}$.
   The reduced phase superspace can be defined by the factorization
   of
   $M_{r^2_0 , 0, 0}$ by  the
  action of the tangent  vector field
$\{{\cal J},\;\}_0$. Hence, the
reduced phase superspace  is a $(2N.2N)_{\DR}$-dimensional one.
The conventional local coordinates
of the reduced phase superspace could be chosen as follows
(on the chart $u^0\neq 0$):
\beq
            \sigma^a=-i\{z^a ,\Theta^+\}=
             \theta^a-\theta^0 z^a,\quad
  w^a= z^a +i\frac{\Theta^-}{{\cal J}_0}\sigma^a,
\eeq
where $z^a, \theta^0,\theta^a$ are defined by (\ref{szt}).
    The reduced Poisson brackets
 are defined as follows:
                    $$
               \{f,g\}_0^{\rm red}=
               \{f,g\}_0\mid_{{\cal J}=r^2_0,\Theta^\pm=0},
                                             $$
      where  $f,g$ are the functions on $(w^a, \sigma^a)$.
 Straightforward calculations result in
  the following  relations:
\begin{eqnarray}
\{w^A,w^B\}_0^{\rm red}&=&\{{\bar w}^A,{\bar w}^B\}_0^{\rm red}= 0 ,\quad {\rm
where}\quad w^A =(w^a ,\sigma^a) \nonumber \\
          \{w^a,{\bar w^b}\}^{\rm red}_0&=&
i\frac{A}{r^2_0}(\delta^{ab}+w^a{\bar w}^b) -
       \frac{\sigma^a{\bar \sigma}^b}{r^2_0},   \nonumber \\
\{ w^a,\bar\sigma^b \}_0^{\rm red}&=&
   i\frac{A}{r^2_0}\left( w^a{\bar \sigma}^b +
     \mu (\delta^{ab}+w^a{\bar w}^b )\right)   \\
 \{\sigma^a,{\bar \sigma^b}\}^{\rm red}_0 &=&
  \frac{A}{r^2_0}\left( (1+i\mu{\bar \mu})\delta^{ab}+w^a{\bar w}^b +
 i(\sigma^a +\mu w^a)({\bar \sigma}^b+{\bar \mu}{\bar w}^b \right),\nonumber
   \end{eqnarray}
 and
$$   A= 1+w^a{\bar w}^a -i\sigma^a {\bar \sigma}^a +
 \frac{i\sigma^a {\bar w^a} {\bar \sigma}^b w^b}{1+w^c{\bar w}^c }, \;\;\;\;
    \mu= \frac{{\bar w}^a \sigma^a}{1+w^b{\bar w}^b } . $$

 These Poisson  brackets are associated with the K\"ahler
  structure defined by the potential
\beq
\begin{array}{c}
   K= r^2_0\log A(w,\bar w, \sigma,\bar\sigma )=
   = r^2_0\log(1+w^a\bar w^a ) +r^2_0\log(1-ig_{a\bar b}\sigma^a\bar\sigma^b).
\end{array}
\eeq
where $g_{a\bar b}(w,\bar w)$ is the Fubini-Study metric on $\DC P^N$.

The transition functions  from the ${\tilde a}$-th chart to the ${\tilde b}$-th one
reads
$$
w^{\tilde c}_{({\tilde b})}=\frac{w^{\tilde c}_{({\tilde a})}}{
w^{\tilde b}_{({\tilde a})}},\quad
\sigma^{\tilde c}_{({\tilde b})}
     = \frac{\sigma^{\tilde c}_{({\tilde a})}x^{\tilde b} _{({\tilde a})}
- w^{\tilde c}_{({\tilde a})}\sigma^{\tilde b}_{({\tilde a})}}
      {(w^{\tilde b}_{({\tilde a})})^2},
      $$
where $(w^{\tilde a}_{({\tilde a})}=1,\sigma^{\tilde b}_{({\tilde b})}=0)$.
Hence, $\sigma^a$ transforms like $dw^a$, i.e.  the reduced phase
 superspace is $\Lambda \DC P^N$, the external algebra
 of the the complex projective space $ \DC P^N$.\\
{\bf Remark 1.} On $\DC ^{N+1,N+1}$
one can define the
  odd K\"ahler structure as well,
  $\Omega^1 = du^n \wedge d{\bar \eta}^n + d{\bar u}^n \wedge d\eta^n $.
It could  be reduced to the odd K\"ahler structure on $\Lambda \DC P^N$
by the action
of the generators
$$
            J_0= z {\bar z},\quad Q = z {\bar \eta} + {\bar z} \eta \;.
$$
{\bf Remark 2.} The generalization of the reduction $T^*\DC^2 \to T^*\DR^3$,
where the latter is specified by the presence of a Dirac monopole,
is also straightforward. One should consider the $(4.M)_{\DC}$-
dimensional superspace equipped with the canonical even symplectic structure
$\Omega_0=d\pi\wedge dz+d\bar\pi\wedge d\bar z +d\eta\wedge d\bar\eta $,
and reduce it by the Hamiltonian action of the $U(1)$ group given by the generator
${\cal J}=i\pi z-i\bar\pi\bar z -i\eta\bar\eta $. The resulting space is a
$(6.2M)_{\DR}$-dimensional one. Its even local coordinates
could be defined by the same expressions, as in the bosonic case, Eq.(\ref{ksc}),
while the odd coordinates could be chosen as follows:
$\theta^m=f(z\bar z){\bar z}_0\eta^m$.

\section{Supersymmetric mechanics}
 In the previous Sections we presented some basic elements of the Hamiltonian and
 super-Hamiltonian formalism.
We paid special attention to the examples, related with K\"ahler geometry, keeping
 in mind
that the latter is of a special importance in supersymmetric mechanics.
Indeed, the incorporation of the K\"ahler structure(s)
is one of the standard ways to increase the number of supersymmetries of the system.

Our goal is to construct the supersymmetric mechanics with ${\cal N}\geq 2$
supersymmetries. This means that, on the given phase superspace equipped with
even symplectic structure, we should construct the Hamiltonian ${\cal H}$
which has ${\cal N}=N$ odd constants of motion $Q_i$ forming the superalgebra
\beq
\{Q_i, Q_j\}=2\delta_{ij}{\cal H}, \qquad \{Q_i,{\cal H}\}=0.
\eeq
This kind of  mechanics is referred to  as ``${\cal N}=N$ supersymmetric mechanics".

 It is very easy to construct the
${\cal N}=1$ supersymmetric mechanics with single supercharges:
we should simply take the square (under a given nondegenerate even Poisson bracket) of the
 arbitrary odd  function $Q_1$, and consider the resulting
 even function as the Hamiltonian
\beq
\{Q_{1},Q_1\}\equiv 2{\cal H}_{SUSY}:\Rightarrow \{Q_1 , {\cal H}_{SUSY}\}=0.
\eeq
However, the case of ${\cal N}=1$ supersymmetric mechanics
 is not an interesting system, both
from the dynamical and field-theoretical viewpoints.

If we want to  construct the ${\cal N}>1$ supersymmetric mechanics,
we must specify  both the underlying system and the structure of phase superspace.

Let us illustrate it on the simplest examples of ${\cal N}=2$ supersymmetric mechanics.
For this purpose, it is convenient to present the ${\cal N}=2$ superalgebra as follows:
\beq
\{Q^+,Q^-\}={\cal H},\quad \{Q^\pm ,Q^\pm \}=0,
\label{2sualgch}\eeq
where $Q^\pm=(Q_1\pm iQ_2)/\sqrt{2}$.
Hence, we have to find the odd complex function,
which  is nilpotent with respect to the given nondegenerate
Poisson bracket, in order to construct the appropriate system.

Let us consider a particular example, when the underlying system is  defined on the
cotangent bundle $T^*M_0$, and it is given by (\ref{12}).

In order to supersymmetrize this system, we extend the canonical
symplectic structure as follows:
\beq
\Omega=dp_a \wedge dx^a +
 \frac 12 R_{abcd}\theta^a_+\theta^b_- dx^c\wedge
dx^d +g_{ab}D\theta^a_+ \wedge D\theta^b_-,
\eeq
where $D\theta^a_\pm\equiv d\theta^a_\pm+
\Gamma^a_{bc}\theta^b_\pm dx^c$,
and    $\Gamma^a_{bc}$, $R_{abcd}$
 are the components of the connection and curvature
 of the metrics
  $g_{ab}dx^a dx^b$ on $M_0$.

We choose the following candidate for a complex supercharge:
\beq
Q_{\pm}=(p_a\pm iW_{,a})\theta^a{_\pm} \;: \{Q_\pm, Q_\pm \}=0.
\eeq
Hence, the supersymmetric Hamiltonian could
 be constructed by the calculation of the Poisson brackets
of these supercharges.
\beq
{\cal H}\equiv\{Q_+, Q_-\}=\frac12 g^{ab}(p_ap_b+W_{,a}W_{,b})
 +W_{a;b}\theta^a_+\theta^b_-
+ R_{abcd}\theta^a_-\theta^b_+\theta^c_-\theta^d_+.
\eeq
The "minimal" coupling of the magnetic field,  $\Omega\to
 \Omega+F_{ab}dx^a\wedge dx^b$, breaks the
    ${\cal N}=2$ supersymmetry of the system
    $$ \{Q_\pm,Q_{\pm}\}=F_{ab}\theta^a_\pm\theta^b_{\pm},
\quad    \{Q_+, Q_-\}= {\cal H}+ iF_{ab}\theta^a_+\theta^b_- .$$
Notice that the Higgs oscillator on the sphere $S^N$, considered in Section 3,
could be supersymmetrised in this way,
choosing
$W=\frac{\alpha}{2} \log\frac{2+{\bf q^2}}{2-{\bf q}^2},$
 with  ${\bf q}$ being the conformal coordinates of the sphere.

One of the ways to extend this construction
 to ${\cal N}=4$ supersymmetric mechanics is
  the doubling of the number of odd degrees of freedom.
 It was considered, within the (Lagrangian) superfield approach in Ref.\cite{dpt}.
In this paper the authors considered the $(2N.2N)_{\DR}$-dimensional
 superspace and  the supercharges containing term cubic on odd variables.
Calculating the Poisson brackets, the authors found
that the admissible metrics of the configuration space of that system
should have the following local form:
\beq
g_{ab}=\frac{\partial^2 A (x)}{\partial x^a\partial x^b}.
\label{pashmet}\eeq
The admissible set of potentials looks, in this local coordinates,
as follows: $V=g_{ab}c^{ab}+g^{ab}d_{af}$, where $c^{ab}$ and $d_{ab}$
are constant matrices.\\

So, considering the Hamiltonian system with generic phase spaces,
we found that without any efforts it could be extended to ${\cal N}=1$
supersymmetric mechanics.
For the construction of ${\cal N}=2$ supersymmetric mechanics
we were forced to restrict ourselves to systems on the
cotangent bundle of Riemann manifolds. Even after this strong restriction,
we found that the inclusion of a magnetic field breaks the supersymmetry
of the system. On the other hand, in trying to construct
${\cal N}=4$ supersymmetric mechanics, we found that in this
case  even the metric of the configuration space and the
admissible set of potentials  are strongly restricted.\\

In further examples we shall show that the transition
to K\"ahler geometry makes these restrictions much weaker.

\subsection*{${\cal N}=2$ supersymmetric mechanics with K\"ahler phase space}
Let us  consider a supersymmetric mechanics
 whose phase superspace is the external
algebra of the K\"ahler manifold $\Lambda{M}$, where
$\left(M,\; g_{a\bar b}(z,\bar z)dz^a d{\bar z}^{\bar b}\right)$
 is the phase space
of the underlying Hamiltonian mechanics \cite{Bellucci:2001dt}.
The phase superspace is
$(D|D)_{\DC}-$ dimensional  supermanifold
 equipped with the K\"ahler
 structure
$$
\Omega=i\partial \bar\partial
\left(K-ig_{a\bar b}\theta^a{\bar\theta}^{\bar b}
\right)
=i(g_{a\bar b}+i
R_{a\bar bc\bar d}\theta^c\bar\theta^{\bar d})dz^a\wedge
 d{\bar z}^{\bar b}
+g_{a\bar b}D\theta^a\wedge D{\bar\theta}^{\bar b},
$$
where
$D\theta^a=d\theta^a +
\Gamma^a_{bc}\theta^c dz^c$, and
 $\Gamma^a_{bc}$, $R_{a\bar bc\bar d}$ are
the Cristoffel symbols and curvature tensor of the underlying
K\"ahler metrics $g_{a\bar b}=\partial_a\partial_{\bar b}K(z,\bar z)$, respectively.

The corresponding Poisson bracket can be presented in the form
\begin{equation}
\{\quad,\quad\}=i{\tilde g}^{a\bar b}\nabla_a
\wedge{\bar\nabla}_{\bar b}+g^{a\bar b}\frac{\partial}{\partial \theta^a}
\wedge\frac{\partial}{\partial{\bar\theta}^{\bar b}}
\end{equation}
where
$$
\nabla_a=\frac{\partial}{\partial z^a}-
\Gamma^c_{ab}\theta^b\frac{\partial}{\partial\theta^c},
\quad
{\tilde g}^{-1}_{a\bar b}=(g_{a\bar b}+
 iR_{a\bar bc\bar d}\theta^c\bar\theta^{\bar d}).
$$
On this phase superspace one can immediately construct
${\cal N}=2$ supersymmetric mechanics,
defined  by the supercharges
\begin{equation}
Q^0_+=\partial_a K(z, \bar z)\theta^a,\quad
Q^0_-=\partial_{\bar a}K(z, \bar z){\bar\theta}^{\bar a}
\label{q0}\end{equation}
where $K(z,\bar z)$
is  the K\"ahler potential of $M$, defined up to holomorphic and
anti-holomorphic functions,
$K(z, \bar z)\to K(z,\bar z)+U(z)+{\bar U}(\bar z)$.

The Hamiltonian of the system reads
\begin{equation}
{{\cal H}}_0=g^{a\bar b}\partial_a K \partial_{\bar b}K
-ig_{a\bar b}\theta^a{\bar\theta}^{\bar b}
+i\theta^cK_{c;a}{\tilde g}^{a\bar b}K_{\bar b;\bar d}{\bar\theta}^{\bar d}
\label{h0}\end{equation}
where
$K_{a;b}=\partial_a\partial_bK-\Gamma^c_{ab}\partial_cK$.
\\
Another example of  ${\cal N}=2$ supersymmetric mechanics
 is defined by the supercharges
\begin{equation}
Q^c_+=\partial_a G(z, \bar z)\theta^a,\quad
Q^c_-=\partial_{\bar a}G(z, \bar z){\bar\theta}^{\bar a},
\label{c0}\end{equation}
where the real function $G(z,\bar z)$
is  the Killing potential of the underlying K\"ahler structure
\beq
 {\partial_a\partial_b}G-\Gamma^c_{ab}\partial_c G=0,\quad
G^a(z)=g^{a\bar b}\partial_{\bar b}G(z,\bar z).
\label{killing}\eeq
In this case the  Hamiltonian of system reads
\begin{equation}
{{\cal H}^c}=g_{a\bar b} G^a G^{\bar b}+
 i{\bar\theta}^{\bar d}G_{a{\bar d}}{\tilde g}^{a\bar b}
G_{c\bar b}\theta^c,
\label{hc}\end{equation}
where $G_{a\bar b}=\partial_a{\partial}_{\bar b}G(z,\bar z)$.

The commutators of the supercharges in  these particular examples
read
\begin{equation}
  \{Q^c_\pm, Q^0_\pm\}={\cal R}_\pm,\quad
\{Q^c_\pm, Q^0_\mp\}={\cal Z},
\end{equation}
where
\begin{equation}
{\tilde{\cal Z}}\equiv G(z,\bar z)+
iG_{a\bar b}(z,\bar z)\theta^a{\bar\theta}^{\bar b},\quad
{\cal R}_+=i\theta^cK_{c;a}{\tilde g}^{a\bar b}G_{\bar b; d}
{\theta}^{d},\quad{\cal R}_-=\bar{\cal R}_+\;\;.
\label{Z}\end{equation}
Hence, introducing the supercharges
\begin{equation}
\Theta_\pm=Q^0_\pm \pm iQ^c_\mp,
\end{equation}
  we can define  $N=2$ SUSY mechanics specified by
the presence of the central charge ${\cal Z}$
\begin{equation}
\begin{array}{c}
\{\Theta_+,\Theta_-\}={\tilde{\cal H}},\quad
 \{\Theta_\pm, \Theta_\pm\}=\pm i{\cal Z}\\
\{{\cal Z},\Theta_\pm\}=0,\;\;-\{{\tilde{\cal H}},\Theta_\mp\}=0,\;\;
\{{\cal Z}, {\tilde{\cal H}}\}=0.
\end{array}
\end{equation}
The Hamiltonian of this
generalized mechanics is defined by the expression
\begin{equation}
{\tilde{\cal H}}={\cal H}_0+{\cal H}_c +i{\cal R}_+ -i{\cal R}_-.
\end{equation}
A ``fermionic number" is of the  form
\begin{equation}
{\tilde {\cal F}}=ig_{a\bar b}\theta^a{\bar\theta}^{\bar b}:
\;
 \{ {\tilde{\cal F}},\Theta_\pm,\}=\pm i \Theta_\pm \; .
\end{equation}
It seems
 that, on the external algebra of the  hyper-Kahler manifold,
 in the same manner one could construct
${\cal N}=4$ supersymmetric mechanics.
On the other hand, the hyper-K\"ahler manifolds
are the cotangent bundle of the K\"ahler manifolds
equipped with Ricci-flat metrics.

We shall demonstrate, in the next examples, that  these
restrictions can be too strong.
 Namely, choosing the underlying phase space to be the
cotangent bundle of the
 K\"ahler manifold, we  will double the number
 of supercharges and get the
  ${\cal N}=4$ supersymmetric mechanics on the cotangent bundles
  of generic K\"ahler manifolds and  the
  ${\cal N}=8$ ones on the cotangent bundles of the special
  K\"ahler manifolds.

\subsection*{${\cal N}=4$ supersymmetric mechanics}
Let us  show that the Hamiltonian  mechanics (\ref{12})
could be easily extended to the ${\cal N}=4$ supersymmetric mechanics,
when the configuration space $M_0$ is the K\"ahler manifold
$(M_0, g_{a\bar b}dz^ad{\bar z}^{\bar b})$,
$g_{a\bar b}=\partial^2 K(z,\bar z)/\partial z^a\partial{\bar z}^b$, and the
potential term
has the form
 $$
 V(z,\bar z)=
 \frac{\partial \bar U(\bar z)}{\partial \bar z^a}g^{\bar ab}
 \frac{\partial U(z)}{\partial z^b}\;.$$
 For this purpose, let us define the supersymplectic structure
\begin{equation}
\begin{array}{c}
\Omega=\omega_0-i\partial{\bar\partial}{\bf g}=\\
=d\pi_a\wedge dz^a+ d{\bar\pi}_a\wedge d{\bar z}^a
+R_{a{\bar b}c\bar d}\eta^a_i\bar\eta^b_i dz^a\wedge d{\bar z}^b+
g_{a\bar b}D\eta^a_i\wedge{D{\bar\eta}^b_i}
\end{array}
\label{ss}\end{equation}
where
$$
{\bf g}=ig_{a\bar b}\eta^a\sigma_0{\bar\eta}^b,\quad
D\eta^a_i=d\eta^a_i+\Gamma^a_{bc}\eta^a_i dz^a,\quad i=1,2
$$
$\Gamma^a_{bc},\; R_{a\bar b c\bar d}$ are
the connection and curvature of the K\"ahler structure, respectively,
and the odd coordinates $\eta^a_i$ belong to the external algebra
$\Lambda M_0$, i. e. they transform as  $dz^a$.
This symplectic structure becomes canonical
in the coordinates $(p_a,\chi^k)$
\begin{equation}
\begin{array}{c}
p_a=\pi_a-\frac{i}{2} \partial_a{\bf g},
\quad\chi^m_i={\rm e}^m_b\eta^b_i\;:\\
\Omega=dp_a\wedge d z^a +d{\bar p}_{\bar a}\wedge d{\bar z}^{\bar a}
+d\chi^m_i\wedge d{\bar\chi}^{\bar m}_i,
\end{array}
\label{canonical}\end{equation}
where ${\rm e}^m_a$ are the einbeins of the  K\"ahler structure:
${\rm e}^m_a\delta_{m\bar m}{\bar{\rm e}}^{\bar m}_{\bar b}=g_{a\bar b}.$
The corresponding Poisson brackets are defined
by the following non-zero
relations (and their  complex-conjugates):
$$
\begin{array}{c}
\{\pi_a, z^b\}=\delta^b_a,\quad
\{\pi_a,\eta^b_i\}=-\Gamma^b_{ac}\eta^c_i,\\
\{\pi_a,\bar\pi_b\}=-R_{a\bar b c\bar d}\eta^c_k{\bar\eta}^d_k,\quad
\{\eta^a_i, \bar\eta^b_j\}=g^{a\bar b}\delta_{ij}.
\end{array}
$$
Let us represent the ${\cal N}=4$ supersymmetry algebra as follows:
 \begin{equation}
\{Q^+_i,Q^-_j\}=\delta_{ij}{\cal H},\quad
\{Q^\pm_i,Q^\pm_j\}=\{Q^\pm_i, {\cal H}\}=0,\quad i=1, 2,
\label{4sualg}\end{equation}
and choose the  supercharges  given by the functions
 \begin{equation}
Q^+_1=\pi_a\eta^a_1+ iU_{\bar a}{\bar \eta}^{\bar a}_2,\quad
Q^+_2=\pi_a\eta^a_2- iU_{\bar a}{\bar \eta}^{\bar a}_1.
\label{4SUSY}\end{equation}
Then, calculating the commutators (Poisson brackets) of these
functions, we get that the supercharges (\ref{4SUSY})  belong
to the superalgebra (\ref{4sualg}),
when the functions $U_a, {\bar U}_{\bar a}$
are  of the form
\begin{equation}
U_a(z)=\frac{\partial U(z)}{\partial z^a},\quad
{\bar U}_{\bar a}(\bar z )=
\frac{\partial {\bar U}({\bar z})}{\partial {\bar z}^a},
\end{equation}
while   the Hamiltonian reads
\begin{equation}
{\cal H}=g^{a{\bar b}}(\pi_a{\bar\pi}_b+
{U}_a{\bar U}_{\bar b}) -iU_{a;b}\eta^a_1\eta^{b}_2
+i{\bar U}_{\bar a;\bar b}{\bar\eta}^{\bar a}_1{\bar\eta}^{\bar b}_2
-R_{a\bar b c\bar d}\eta^a_1\bar\eta^b_1\eta^a_2\bar\eta^d_2,
\label{4SUHam}\end{equation}
 where
$ U_{a;b}\equiv \partial_a\partial_b U-\Gamma^c_{ab}\partial_cU$.
%
%

Now, following \cite{n4}, let us extend this system to
 ${\cal N}=4$ supersymmetric mechanics with central charge
\begin{equation}
\{\Theta^+_i,\Theta^-_j\}=
\delta_{ij}{\cal H}+{\cal Z}\sigma^3_{i{j}},\quad
\{\Theta^\pm_i,\Theta^\pm_j\}=0,
\{{\cal Z}, {\cal H}\}= \{{\cal Z},\Theta^\pm_k\}=0.
\label{csa}\end{equation}
For this purpose one introduces
the supercharges
\begin{equation}
\begin{array}{c}
 \Theta^+_1=\left(\pi_a+iG_{,a}(z,\bar z)\right)\eta^a_1 +
i {\bar U}_{,\bar a}({\bar z}){\bar \eta}^{\bar a}_2,\\
 \Theta^+_2=\left(\pi_a-iG_{,a}(z,\bar z)\right)\eta^a_2 -
i {\bar U}_{,{\bar a}}({\bar z}){\bar \eta}^{\bar a}_1,
\end{array}
\end{equation}
where the real function $G(z,\bar z)$ obeys the
 conditions (\ref{killing})
 and
 $\partial_{\bar a}G g^{\bar a b}U_b=0 $.
So, $G$ is a Killing potential
defining the isometry of the underlying K\"ahler manifold
 (given by the vector $
{\bf G}=G^a(z)\partial_a+{\bar G}^a({\bar z}){\bar\partial}_a,\quad
G^a=ig^{a\bar b}{\bar \partial}_b G$)
which leaves the holomorphic function
$U(z)$  invariant
$${\cal L}_{\bf G}U=0\;\Rightarrow \;G^a(z)U_a(z)=0.$$

Calculating the Poisson brackets of these supercharges,
we get explicit
expressions for the Hamiltonian
\begin{equation}
\begin{array}{c}
{\cal H}\equiv
g^{a{\bar b}}\left(\pi_a{\bar\pi}_{\bar b}+ G_{,a}G_{{\bar b}}
+{ U}_{,a}{\bar U}_{,\bar b}\right)- \\
-iU_{a;b}\eta^a_1\eta^{b}_2 +
i{\bar U}_{\bar a;\bar b}{\bar\eta}^{\bar a}_1{\bar\eta}^{\bar b}_2
+\frac 12 G_{a\bar b}(\eta^a_k\bar\eta^{\bar b}_k)
-R_{a\bar b c\bar d}\eta^a_1\bar\eta^b_1\eta^c_2\bar\eta^d_2
\end{array}
\end{equation}
and for the central charge
\begin{equation}
{{\cal Z}}=i(G^a\pi_a+G^{\bar a}{\bar\pi}_{\bar a})+\frac{i}{2}
\partial_a{\bar\partial}_{\bar b}G(\eta^a{\sigma_3}\bar\eta^{\bar b}).
\label{z}\end{equation}
It can be checked by a straightforward calculation that
the function ${\cal Z}$ indeed belongs to the center of
the superalgebra (\ref{csa}).
The scalar part of each  phase
with  standard ${\cal N}=2$ supersymmetry can be interpreted
as a particle  moving on the K\"ahler  manifold
in the presence of an external magnetic field, with strength
$F=iG_{a\bar b}dz^a\wedge d{\bar z}^{\bar b}$, and in the
potential field $U_{,a}(z)g^{a\bar b}{\bar U}_{,\bar b}(\bar z)$.

Assuming that $(M_0, g_{a\bar b}dz^a d{\bar z}^b)$ is the
hyper-K\"ahler metric, $U(z)+{\bar U}({\bar z})$
is a tri-holomorphic function and
$G(z,\bar z)$ defines a tri-holomorphic Killing vector,
one should get ${\cal N}=8$ supersymmetric mechanics.
In this case, instead of the
phase with standard ${\cal N}=2$ supersymmetry arising in the
K\"ahler case, we shall get the phase with standard
${\cal N}=4$ supersymmetry.
%
This system could be straightforwardly constructed
 by the dimensional reduction of the ${\cal N}=2$ supersymmetric
$(1+1)$ dimensional sigma-model by
 Alvarez-Gaum\'e and Freedman \cite{agf}.

\subsubsection*{${\cal N}=8$ mechanics}
We have seen
 that the transition from the generic Riemann space to the generic K\"ahler space
 allows one to double the number of supersymmetries from
  ${\cal N}=2$ to ${\cal N}=4$,
 with the appropriate restriction of the admissible set of potentials.

On the other hand, we mentioned that the
doubling of the number of odd variables and the restriction
the Riemann metric allow one to construct the ${\cal N}=4$ supersymmetric mechanics
\cite{dpt}.
Now, following the paper \cite{n8}, we shall show that a similar procedure,
applied to the systems on K\"ahler manifolds, permits to construct the ${\cal N}=8$
supersymmetric mechanics, with the supersymmetry algebra
\footnote{We use the following
convention: $\epsilon_{ij}\epsilon^{jk}=\delta_i^k,\;
\epsilon_{12}=\epsilon^{21}=1$ .} \beq
\{Q_{i\alpha},Q_{j\beta}\}=\{\bQ_{i\alpha},\bQ_{j\beta}\}= 0,\quad
\{Q_{i\alpha},\bQ_{j\beta}\}=
\epsilon_{\alpha\beta}\epsilon_{ij}{\cal H}_{SUSY},\label{188}\eeq
where $i,j\;=\; 1,2$, $\alpha ,\beta =1, 2$.

 We present the results for the mechanics without (bosonic)
potential term. The respective systems with potential terms are constructed
in \cite{bks}.

In order to construct the ${\cal N}=8$ supersymmetric mechanics,
let us
define the   $(2d.4d)_{\DC}$-dimensional symplectic structure
$$
\Omega=d{\cal A}=d\pi_a\wedge dz^a+ d{\bar\pi}_a\wedge d{\bar z}^a
- R_{a\bar bc\bar d}\eta^c_{i\alpha}\bar\eta^{d|i\alpha}
dz^a\wedge d{\bar z}^b+ g_{a\bar b}D\eta^a_{i\alpha} \wedge
D{\bar\eta}^{b| i\alpha}\quad ,$$
 where
\beq\label{a}
 {\cal A}=\pi_adz^a+\bar\pi_ad{\bar z}^a
+\frac 12\eta^a_{i\alpha}g_{a\bar b}D\bar\eta^{b|i\alpha}+ \frac
12\bar\eta^b_{i\alpha}g_{a\bar b}D\eta^{a| i\alpha},
\eeq
 and $D\eta^a_{i\alpha} =d\eta^a_{i\alpha}+\Gamma^a_{bc}\eta^b_{i\alpha}
dz^c$.
The corresponding Poisson brackets are given by the
following non-zero relations (and their complex-conjugates):
\beq
\begin{array}{c}
\{\pi_a, z^b\}=\delta^b_a,\quad
\{\pi_a,\eta^b_{i\alpha}\}=-\Gamma^b_{ac}\eta^c_{i\alpha},\\
\{\pi_a,\bar\pi_b\}= R_{a\bar bc\bar d}\eta^c_{i\alpha}{\bar\eta}^{d|i\alpha},
 \quad \{\eta^a_{i\alpha},
\bar\eta^{b|j\beta}\}= g^{ab}\delta_i^j\delta_\alpha^\beta.
\end{array}
\eeq
Let us search the supercharges among the functions
\beq
Q_{i\alpha}=\pi_a\eta^a_{i\alpha}+\frac 13 {\bar f}_{abc}
{\bar T}^{abc}_{i\alpha},
\quad
\bQ_{i\alpha}={\bar\pi}_a{\bar\eta}^a_{i\alpha}+
\frac 13 f_{abc}T^{abc}_{i\alpha}\label{qqd}\eeq
 where $ T^{abc}_{i\alpha}\equiv
\eta^a_{i\beta}\eta^{bj\beta}\eta^c_{j\alpha}$.

Calculating  the mutual Poisson brackets of
$Q_{i\alpha},\bQ_{i\alpha}$
 one can get, that they obey the
 ${\cal N}=8$ supersymmetry algebra, provided the following relations
 hold:
\beq \frac{\partial}{\partial{\bar z}^d}f_{abc}=0\; ,
\qquad R_{a\bar b c\bar d}=-
f_{ace}g^{ee'}{\bar f}_{e'bd}. \label{comp}\eeq The above
equations  guarantee, respectively, that  the first and
second equations in (\ref{188}) are fulfilled.
 Then we could immediately get the
  ${\cal N}=8$  supersymmetric  Hamiltonian
\beq
{\cal H}_{SUSY}=
\pi_a g^{ab}\bar\pi_b+ \frac 13 f_{abc;d}\Lambda^{abc d} +\frac
13{\bar f}_{abc;d}\bar\Lambda^{abcd} + f_{abc}g^{c\bar c'}{\bar f}_{c'
de} \Lambda_0^{ab\bar d\bar e} ,\eeq
where
$$\Lambda^{abcd}\equiv -\frac 14
\eta^a_{i\alpha}\eta^{bi\beta}\eta^c_{k\beta}\eta^{dk\alpha},\quad
\Lambda_0^{ab\bar c\bar d}\equiv\frac 12
(\eta^{a\alpha}_i\eta^b_{j\alpha}\bar\eta^{c\beta i}
\bar\eta^{dj}_\beta + \eta^{ai}_\alpha\eta^b_{i\beta}
\bar\eta^{cj\alpha} \bar\eta^{d\beta}_{j}),$$
and
$f_{abc;d}=f_{abc,d}-
\Gamma^e_{da}f_{ebc}-\Gamma^e_{db}f_{aec}-\Gamma^e_{dc}f_{abe}
$ is the
covariant derivative of the third rank covariant symmetric tensor.\\
The equations (\ref{comp})  precisely  mean that the configuration space
$M_0$ is a {\sl special K\"ahler manifold  of the rigid type} \cite{fre}.
Taking into account the symmetrizing of $f_{abc}$ over spatial indices and
the
explicit expression of $R_{a\bar b c \bar d}$ in terms of the metric $g_{a b}$,
we can  immediately  find the local solution for equations (\ref{comp})
\beq
f_{abc}=
\frac{\partial^3 f(z)}{\partial z^a\partial z^b\partial z^c},\qquad
g_{a\bar b}={\rm e}^{i\nu}\frac{\partial^2 f(z)}{\partial z^a\partial z^b}
+
{\rm e}^{-i\nu}
\frac{\partial^2 \bar f(\bar z)}{\partial \bar z^a\partial \bar z^b},
\label{gsk}
\eeq
where $\nu=const\in \DR$.\\
Redefining the local function, $f\to i{\rm e}^{-i\nu}f$,
we shall get the $\nu$-parametric family of supersymmetric
mechanics, whose metric is defined by the
K\"ahler potential  of a special
K\"ahler manifold of the rigid type.
Surely, this local solution is not covariant under arbitrary holomorphic
transformation, and it assumes the choice of a distinguished
 coordinate frame.

The special K\"ahler manifolds of the rigid type
 became widely known  during last decade due to the so-called
``T-duality symmetry": in the context of ${\cal N}=2,
 d=4$ super Yang-Mills theory,  it
 connects the UV and IR limit of the theory\cite{SW}.
The ``T-duality symmetry''
 is expressed in the line below
\beq
\left( z^a,\; f(z)\right)\Rightarrow
\left( u_a=\frac{\partial f(z)}{\partial z^a},\; {\widetilde f}(u)\right),
\quad {\rm where}\quad
 \frac{\partial^2{\widetilde f}(u)}{\partial u_a\partial u_c}
\frac{\partial f}{\partial z^c\partial z^b}=-\delta^a_b\; .
\label{duality}\eeq
It is clear that the symplectic structure is {\sl covariant} under
the following holomorphic transformations:
\beq
{\tilde z}^a= {\tilde z}^a(z),\quad {\tilde\eta}^a_{i\alpha}=\frac{\partial
  {\tilde z}^a(z)}{\partial z^b}\eta^b_{i\alpha},
\quad{\tilde\pi}_a=\frac{\partial z^b}{\partial{\tilde z}^a}\pi_b,
\label{ht}\eeq
By the use of (\ref{ht}), we can extend the duality transformation
(\ref{duality})  to  the
whole phase superspace $
(\pi_a,\,  z^a ,\, \eta^a_{i\alpha})\to (p^a, u_a, \psi_{a|i\alpha})$
\beq
 {u}_a=\partial_a f(z),\quad
p^a\frac{\partial^2 f}{\partial z^a\partial z^b}=-\pi_b
 ,\quad \psi^{i\alpha}_a=
\frac{\partial^2 f}{\partial z^a\partial z^b}\eta^{b|i\alpha}.
\eeq
Taking into account the expression of the symplectic structure in terms of the
presymplectic one-form (\ref{a}), we can easily perform the Legendre
transformation of the Hamiltonian to the (second-order) Lagrangian
\bea
&&{\cal L}= {\cal A}(d/dt)-{\cal H}_{SUSY}\vert_{\pi_a=g_{a\bar b}
{\dot{\bar z}}^b}
= \nonumber\\
&&=g_{a\bar b}{\dot z}^a{\dot {\bar z}}^b
+\frac 12\eta^a_{i\alpha}g_{a\bar b}\frac{D\bar\eta^{b|i\alpha}}{dt}+
\frac 12 \bar\eta^b_{i\alpha}g_{a\bar  b}\frac{D\eta^{a|i\alpha}}{dt}-
\label{lag}\eea
$$
-\frac 13
f_{abc;d}\Lambda^{abcd} -\frac 13{\bar f}_{abc;d}\bar\Lambda^{abcd} -
f_{abc}g^{c\bar c'}{\bar f}_{c' de}\Lambda_0^{ab\bar d\bar e}.
$$
Here we denoted $d/dt={\dot z}^a\partial/\partial z^a+
\dot\eta^a_{i\alpha}\partial/\partial\eta^a_{i\alpha} +c.\; c.\;$.
Clearly, the  Lagrangian (\ref{lag})
is covariant under holomorphic transformations (\ref{ht}),
 and the duality transformation as well.
The prepotential ${\widetilde f}(u)$ is connected with $f(z)$ by the
Legendre transformation
$$
{\widetilde f}(u)={\widetilde f}(u,z)\vert_{u_a=\partial_a{f}(z)},
\qquad {\widetilde f}(u,z)=u_az^a-f(z).
$$

\subsection*{Supersymmetric K\"ahler Oscillator}
So far, the K\"ahler structure allowed us to double  the number of supersymmetries
in the system. One can hope that in some cases this could be
 preserved after inclusion of constant magnetic field,
 since this field usually respects the K\"ahler structure.
We shall show,  on the example of the K\"ahler oscillator (\ref{Koscg}),
that it is indeed a case.

Let us consider, following \cite{cpn,sqs},  the supersymmetrization of a specific model of
Hamiltonian mechanics  on the K\"ahler manifold
$(M_0, g_{a\bar b}dz^ad{\bar z}^{\bar b})$
interacting with the constant magnetic field $B$, viz
\beq
{\cal H}=g^{a\bar b}(\pi_a{\bar \pi}_b+
\alpha^2\partial_a K {\bar\partial}_b K),\quad \Omega_0=
d\pi_a\wedge dz^a+ d{\bar\pi}_a\wedge d{\bar z}^a
+iBg_{a{\bar b}} dz^a\wedge d{\bar z}^b,
\label{0}\eeq
where $K(z,\bar z)$ is a K\"ahler potential of configuration space.

Remind,  that the K\"ahler potential  is defined up to holomorphic and antiholomorphic terms,
$K\to K+ U(z)+ \bar U (\bar z)$. Hence,
in the limit $\omega\to 0$  the above Hamiltonian
 takes the form
\beq
{\cal H}=g^{a\bar b}(\pi_a{\bar \pi}_b+
\partial_a U(z) {\bar\partial}_b \bar U(\bar z)),
\eeq
i.e.it  admits, in the absence of magnetic field,
 a ${\cal N}=4$ superextension.

Notice, also, that in the  ``large mass limit", $\pi_a\to 0$,
this system results
in
 the following one:
$${\cal H}_0=\omega^2 g^{a\bar b}\partial_a K {\bar\partial}_b K,\quad
\Omega_0= iBg_{a\bar b}dz^a\wedge d\bar z^b,$$
which could be easily extended to  ${\cal N}=2$ supersymmetric mechanics.

We shall show that, although the system under
 consideration does not possess a standard ${\cal N}=4$
 superextension, it admits  a superextension in terms of
 a nonstandard superalgebra with four fermionic generators,
 including, as subalgebras, two copies of the ${\cal N}=2$ superalgebra.
 This nonstandard superextension
 respects the inclusion of a constant magnetic field.

We  use the following strategy.
 At first, we extend the initial phase space to
 a $(2N.2N)_{\DC}$-dimensional  superspace
 equipped with the symplectic structure
\begin{equation}
\Omega=\Omega_B-iR_{a{\bar b}c\bar d}\eta^c_i\bar\eta^d_i)
dz^a\wedge d{\bar z}^b+ g_{a\bar
b}D\eta^a_i\wedge{D{\bar\eta}^b_i}\quad ,
\end{equation} where $\Omega_B$ is given by (\ref{ssB}).
The corresponding Poisson brackets are defined by the
following non-zero relations (and their complex-conjugates):
\beq
\begin{array}{c}
\{\pi_a, z^b\}=\delta^b_a,\quad
\{\pi_a,\eta^b_i\}=-\Gamma^b_{ac}\eta^c_i ,\\
\{\pi_a,\bar\pi_b\}=i(Bg_{a\bar b}+ i R_{a\bar b c\bar
d}\eta^c_i{\bar\eta}^d_i), \quad \{\eta^a_i,
\bar\eta^b_j\}= g^{a\bar b}\delta_{ij}.
\end{array}
\eeq
Then, in order to  construct the system with the exact  ${\cal N}=2$
supersymmetry (\ref{2sualgch}),
we shall search for the odd functions $Q^\pm$, which obey the equations
$\{Q^\pm,Q^\pm\}=0$ (we  restrict ourselves to the supersymmetric mechanics
whose  supercharges  are {\it linear}  in the
Grassmann variables $\eta^a_i$, $\bar\eta^{\bar a}_i$).

 Let us search for the realization of supercharges
among the functions
\beq Q^\pm=\cos\lambda\;\Theta^\pm_1
+\sin\lambda\;\Theta^\pm_2\;, \eeq where \beq \Theta^+_1=\pi_a
\eta^a_1+ i\bar\partial_a W {\bar \eta}^a_2,\quad
\Theta^+_2={\bar\pi}_a\bar\eta^a_2 +i\;\partial_a W \eta^a_1,\quad
\quad \Theta^-_{1,2}={\bar\Theta}^+_{1,2}, \eeq
and $\lambda$ is
some parameter.

 Calculating the Poisson brackets of the functions,
we get
\begin{eqnarray}
&\{Q^\pm,Q^\pm\}=& i(B\sin 2\lambda\; +
2\alpha\cos 2\lambda ){\cal F}_\pm ,\label{pp}\\
&  \{Q^+,Q^-\}=& {\cal H}^0_{SUSY}+ \left(B\cos 2\lambda\;
-2\alpha\sin 2\lambda \right)\;{\cal F}_3/2. \label{pm}\end{eqnarray}
 Here and further, we use the notation \beq {\cal H}^0_{SUSY}={\cal H}
-R_{a\bar b c\bar d}\eta^a_1\bar\eta^b_1\eta^c_2\bar\eta^d_2
-iW_{a;b}\eta^a_1\eta^b_2+ iW_{\bar a;\bar
b}\bar\eta^a_1\bar\eta^b_2+
 B \frac{{ig_{a\bar b}\eta^a_i{\bar\eta}^b_i}}{2}, \label{hosup}
\eeq
where  ${\cal H}$ denotes the oscillator Hamiltonian (\ref{Koscg}),
and
\beq
{\bf F}= \frac{i}{2}g_{a\bar b}\eta^a_i\bar\eta^b_j\bs_{ij},
\quad{\cal F}_\pm={ F}_1\pm{ F}_2.
 \eeq
One has, then
\beq \{Q^\pm,
Q^\pm\}=0\Leftrightarrow B\sin\;2\lambda+2\alpha\cos\;2\lambda=0,
\eeq so that
$ \lambda=\lambda_0+(i-1)\pi/2$, $i=1,2.$\\
 Here the parameter   $\lambda_0$ is defined by the expressions\beq \cos
2\lambda_0=\frac{B/2}{\sqrt{\alpha^2+(B/2)^2}},\quad \sin
2\lambda_0=-\frac{\alpha}{\sqrt{\alpha^2+(B/2)^2}}. \eeq
Hence, we get the following supercharges: \beq
 Q^\pm_\nu=\cos\lambda_0\Theta^\pm_1+
(-1)^\nu\sin\lambda_0\Theta^\pm_2, \eeq
 and  the pair of
 ${\cal N}=2$ supersymmetric Hamiltonians
\beq
{\cal H}^i_{SUSY}= \{Q^+_\nu ,Q^-_\nu\}=   {\cal H}^0_{SUSY}
-(-1)^i
{\sqrt{\alpha^2+(B/2)^2}} {\cal F}_3 .\label{n4}\eeq
Notice that the supersymmetry invariance is preserved in the presence of a
constant magnetic field.

Calculating the commutators of $Q^\pm_1$ and  $Q^\pm_2$, we get
\beq \{Q^\pm_1,Q^\pm_2\}=2{\sqrt{\alpha^2+(B/2)^2}} {\cal F}_\pm , \quad
\{Q^+_1,Q^-_2\}=0.
\eeq The Poisson brackets between ${\cal F}_\pm $ and
$Q^\pm_\nu$ look as follows: \beq
\begin{array}{c}
\{Q^\pm_i, {\cal F}_\pm\}=0,\quad \{Q^\pm_i, {\cal
F}_\mp\}= \pm\epsilon_{ij}Q^\pm_j,\quad
\{Q^\pm_i, {\cal F}_3\}=\pm iQ^\pm_i .
\end{array}
\eeq
In the notation
$
S^\pm_1\equiv Q^\pm_1,\quad S^\pm_2=Q^\mp_2
$
the whole superalgebra reads
\beq
\begin{array}{c}
\{S^\pm_i, S^\mp_j\}=\delta_{ij}{\cal H}^0_{SUSY}+
\Lambda\sigma^\mu_{ij}{\cal F}_\mu,\\
\{S^\pm_i, {\cal F}_\mu\}=\pm\imath\sigma^\mu_{ij}S^\pm_{j}, \quad
\{{\cal F}_\mu, {\cal F}_{\nu}\}=\epsilon_{\mu\nu\rho}{\cal F}_\rho,
\end{array}
\eeq
where
\beq
\Lambda={\sqrt{\omega^2+(B/2)^2}}.
\eeq
This is precisely the weak supersymmetry algebra  considered by A. Smilga \cite{smilga}.
In the particular case $\omega=0$
it yields the ${\cal N}=4$ supersymmetric mechanics broken
by the presence of a constant magnetic field.
\\

Let us notice  the $\alpha$ and $B$ appear in this superalgebra
in a symmetric way,
via the factor ${\sqrt{\alpha^2+(B/2)^2}}$.

{\bf Remark} In the case of the oscillator on $\DC^N$ we can smoothly
relate the above supersymmetric oscillator  with a ${\it N=4}$ oscillator,
provided we choose
$$
K=\cos{\gamma}\; z\bar z+\sin{\gamma}\;( z^2+ \bar z^2)/2
\; ,\qquad \gamma \in [0,\pi/2].
$$
Hence,
$$
{\cal H}=\pi\bar\pi+\alpha^2_0 z\bar z
+\sin2\gamma\; \alpha^2_0 (z^2+\bar z^2)/2\;,
$$
i.e. for $\gamma=0,\pi /2$ we have a standard harmonic oscillator, while
for $\gamma\neq 0, \pi/2$ we get the anisotropic one, which is
 equivalent to two sets of $N$ one-dimensional oscillators
with frequencies  $\alpha_0{\sqrt{1\pm\sin2\gamma}}$.
The frequency $\alpha$ appearing in the superalgebra,
is of the form: $\alpha=\alpha_0\cos\gamma $.

 \section*{Conclusion}
 We presented some constructions of the Hamiltonian formalism related with Hopf maps
 and K\"ahler geometry, and a few models of supersymmetric mechanics on
 K\"ahler manifolds.
One can hope that the former constructions could be useful in supersymmetric mechanics
along the following lines. Firstly, one could try to extend the number
of supersymmetries, passing from the K\"ahler manifolds to quaternionic
ones. The model suggested in \cite{sutulin} indicates that this could indeed work.
One could also expect that the latter system will respect
the inclusion of an instanton field.
Secondly, one can try to  construct the supersymmetric mechanics,
performing the Hamiltonian reduction of the existing systems, related with the Hopf maps.
In this way one could get new supersymmetric models,
specified by the presence of Dirac and Yang monopoles,
as well as with constant magnetic and instanton fields.\\

{\large Acknowledgements} I would like to thank Stefano Bellucci
for  organizing the Winter School on {\sl Modern trends in
supersymmetric mechanics} and inviting me to deliver (and to write
down) these lectures, which are the result of extensive
discussions with  him, and with S.Krivonos and V.P.Nair. Most of
the examples included in the paper were obtained in
collaboration. I am indebted to all of my co-authors, especially
 to
 A.Yeranian, O.Khudaverdian, V.Ter-Antonyan,
P.-Y.Casteill.

Special thanks to Stefano Bellucci for careful reading of
manuscript and substantial improving of the English.

\end{document}